\documentclass{aa}  
\usepackage{graphicx}
\usepackage[varg]{txfonts}
\usepackage{natbib,twoopt}
\usepackage[breaklinks=true]{hyperref} 
\hypersetup{
    colorlinks=true,
    linkcolor=blue,
    citecolor=blue,
    filecolor=magenta,      
    urlcolor=cyan,
    pdftitle={Overleaf Example},
    pdfpagemode=FullScreen,
    }
\bibpunct{(}{)}{;}{a}{}{,}             

\makeatletter
  \newcommandtwoopt{\citeads}[3][][]{\href{http://adsabs.harvard.edu/abs/#3}%
    {\def\hyper@linkstart##1##2{}%
     \let\hyper@linkend\@empty\citealp[#1][#2]{#3}}}
  \newcommandtwoopt{\citepads}[3][][]{\href{http://adsabs.harvard.edu/abs/#3}%
    {\def\hyper@linkstart##1##2{}%
     \let\hyper@linkend\@empty\citep[#1][#2]{#3}}}
  \newcommandtwoopt{\citetads}[3][][]{\href{http://adsabs.harvard.edu/abs/#3}%
    {\def\hyper@linkstart##1##2{}%
     \let\hyper@linkend\@empty\citet[#1][#2]{#3}}}
  \newcommandtwoopt{\citeyearads}[3][][]%
    {\href{http://adsabs.harvard.edu/abs/#3}
    {\def\hyper@linkstart##1##2{}%
     \let\hyper@linkend\@empty\citeyear[#1][#2]{#3}}}
\makeatother

\usepackage{graphicx}	
\usepackage{color}
\usepackage{array}
\usepackage{multirow}
\usepackage{makecell}
\usepackage{ae,aecompl}
\usepackage[utf8]{inputenc}
\usepackage{url}

\definecolor{darkgreen}{rgb}{0.0,0.5,0.0}
\definecolor{darkred}{rgb}{0.75,0.,0.2}
\definecolor{magenta}{rgb}{0.8,0,0.8}
\definecolor{purple}{rgb}{0.5,0,0.5}
\definecolor{gray}{rgb}{0.5,0.6,0.7}

\newcommand\boldblue[1]{\textcolor{blue}{\mathbf{#1}}}
\newcommand\boldred[1]{\textcolor{darkred}{\mathbf{#1}}}

\newcommand{\diazg}{D\'{\i}az-Gim\'enez}

\newcommand{\CB}{Control\textsubscript{4B}} 
\newcommand{\CC}{Control\textsubscript{4C}} 
\newcommand{\CG}{CG\textsubscript{4}} 
\newcommand{\PCF}{RG\textsubscript{4}} 
\newcommand{\kms}{\,\mathrm{km\,s}^{-1}}

\newcommand{\msun}{\mathcal{M}_\odot}
\newcommand{\LT}{Lim-Tempel}
\newcommand{\LTzMr}{LT$_{z,Mr}$}
\newcommand{\centercell}[1]{\multicolumn{1}{c}{#1}}
\newcommand{\head}[1]{\centercell{\bfseries#1}}

\defcitealias{DiazGimenez+12}{Euge+12}
\defcitealias{DiazGimenez+18}{Euge+18}
\defcitealias{DiazGimenez+21}{Euge+21}

\begin{document} 

\title{Are  compact groups of galaxies special?}

   \author{Matthieu Tricottet\inst{1} \thanks{matthieu.tricottet@gmail.com}
           \and Gary A. Mamon\inst{2}
           \and Eugenia \diazg\inst{3,4}}

   \institute{  
                17, rue des plantes, 75014 Paris, France
                \and Institut d'Astrophysique de Paris (UMR 7095: CNRS \& Sorbonne Universit\'e), 98 bis boulevard Arago, 75014 Paris, France
                \and CONICET. Instituto de Astronom\'ia Te\'orica y Experimental (IATE), Laprida 854, X5000BGR, C\'ordoba, Argentina
                \and Universidad Nacional de C\'ordoba (UNC). Observatorio Astron\'omico de C\'ordoba (OAC), Laprida 854, X5000BGR, C\'ordoba, Argentina
            }

   \date{Received 29 July 2024 /  Accepted  23 April 2025}

  \abstract{
   It is often believed that isolated compact groups (CGs) of galaxies are special systems,
   but only a few studies have compared CGs to regular groups.
   We study the global properties and internal correlations of a volume- and luminosity-complete subsample of 78 groups of four members (\CG{}s) within the HMCG Hickson-like sample of compact groups. We compared these CGs to those of a similarly built subsample (including the three-magnitude range of \CG{s}) of the Lim regular groups. The latter were split into three control samples:
one with the four brightest members (\CB s), one with the four closest members to the brightest group galaxy (BGG; \CC s), and one with exactly four members (\PCF s).
   The vast majority of the \CG{}s are located within regular groups, and a large preponderance of the BGGs of these \CG{}s are the same as those of their host groups.
The \CG{}s are smaller than the groups of all other samples and more luminous than \PCF s. Both results are a consequence of their selection as high surface brightness systems. However, the \CG{}s (especially those split among several regular groups) have luminosities similar to \CC s.
The \CG{}s also have higher velocity dispersions, probably because of a too-permissive redshift accordance criterion.
The BGGs of the \CG{}s are not any more dominant in luminosity than those of \PCF{}s, but they are significantly more offset relative to the group size because the Lim groups are built around their BGGs.
   In summary, compact groups have similar properties to regular groups of four galaxies and to the cores of regular groups once the selection criteria of CGs are considered. A large fraction of the CGs are the cores of regular groups, which are isolated on the sky by construction but rarely isolated in real space (from simulations), indicating that they are often plagued by chance alignments of host group galaxies along the line of sight.
   }


   \maketitle
\nolinenumbers

\section{Introduction}\label{sec:intro}
    
Isolated compact groups (CGs) of four or more galaxies of comparable luminosity, though fairly rare, constitute a potentially extremely dense environment that serves as a unique laboratory to study galaxy interactions. The natural point of view is that CGs constitute physically dense systems \citep{Hickson&Rood88}. But many authors have challenged this view by suggesting that CGs are unbound \citep{Burbidge&Burbidge61}, transient (\citealt{Rose77} for elongated CGs), or mainly contaminated by chance alignments of galaxies along the line of sight \citep{Mamon86,Walke&Mamon89}.
Semi-analytical models (SAMs) of galaxy formation have been used to test the chance alignment hypothesis, starting with the works of \cite*{McConnachie+08} and \cite{DiazGimenez&Mamon10}. The latest, \citep{DiazGimenez+20}, indicates that roughly half of Hickson-like samples in SAMs are indeed chance alignments. This fraction is higher for a greater cosmological density parameter, $\Omega_\mathrm{m}$, and for a lower amplitude of the power spectrum of primordial density fluctuations, $\sigma_8$, but is also lower with higher resolution of the parent dark matter only cosmological simulation on which the SAM was run.

If CGs are indeed extremely dense non-transient environments, one may wonder whether their global properties differ from those of other groups.
This question is best answered with well-defined catalogs of CGs. The most popular of such catalogs was assembled by \cite{Hickson82}, who first searched on photographic plates for isolated compact groups of at least four galaxies within a three-magnitude range. His compactness criterion was based on mean surface brightness, and it led to the HCG catalog of exactly 100 groups. Later, a spectroscopic follow-up \citep{Hickson+92} indicated that 69 of the groups contained at least four galaxies of concordant redshifts. 
The HCG catalog suffers from not only its small size but also from biases. The most important of these biases is the one against CGs dominated by a single galaxy (but still satisfying the three-magnitude range), as found from automatic CG selection from photographic plates \citep*{Prandoni+94} and from expectations from semi-analytical models \citep{DiazGimenez&Mamon10}.

Several catalogs of CGs have been produced by automatic selection, most with highly incomplete photometric or spectroscopic data. The notable exceptions are those extracted from the 
 Two Micron All-Sky Survey (2MASS) Redshift Survey
 (i.e., \citealp{DiazGimenez+12}) 
 and from the Sloan Digital Sky Survey
 (SDSS; i.e, \citealt{McConnachie+09,Sohn+15,Sohn+16}; \citealt*{DiazGimenez+18,Zheng&Shen20,Zandivarez+22}).
 While \cite{McConnachie+09} managed to produce two large catalogs of CGs (with nearly 2300 and 75\,000 groups), only two CGs within them had complete redshift coverage and satisfied the HCG criteria of surface brightness and accordant redshifts \citep{DiazGimenez+18}. 
Similarly, the catalogs of \cite{Sohn+15,Sohn+16} contain 300 and nearly 1600 CGs, among which only 11 and 142 respectively meet the HCG criteria. 

One of the largest catalogs of HCGs with complete spectroscopy and meeting the HCG criteria is the Hickson Modified Compact Group (HMCG) sample of \cite{DiazGimenez+18}, who extracted groups from the SDSS Main Galaxy Sample, yielding 406 CGs with good flags. \cite{DiazGimenez+18} obtained this large sample by performing their CG selection in one step instead of the original two-phase selection (i.e., first photometric, then spectroscopic). Indeed, photometric selection discards groups that seem not to be isolated, whereas their neighbors are often later found to have discordant redshifts and are thus unrelated.
\cite{Zheng&Shen20} built a sample (``cCG") of 6144 CGs of at least three members that meet the HCG criteria and have complete redshift information.
Their cCG sample is mainly based on the SDSS Main Galaxy Sample, but it is supplemented with redshifts from the Large sky Area Multi-Object fiber Spectroscopic Telescope (LAMOST)  \citep{Luo+15} and the Galaxy Mass Assembly (GAMA) \citep{Liske+15} surveys, and as well as from other sources. It contains 847 groups of over four galaxies and 698 groups with exactly four galaxies. It is the largest compact group sample with complete redshift information.
However, the \cite{Zheng&Shen20} catalog includes groups with magnitude ranges greater than three magnitudes as well as many more groups with too-faint brightest galaxy magnitudes to allow for a possible range of three magnitudes, which is important for studying luminosity functions and magnitude gaps. Their sample contains only 135 groups of four galaxies with an SDSS red magnitude  $r_\mathrm{bright}\leq 14.77$, while none have more than four members.   
Finally, \cite{Zandivarez+22} produced a catalog with 1412 CGs of at least three members, but it contains only 300 CGs with four or more galaxies.

A fundamental question concerning CGs is how their properties compare to those of ``regular" groups. This comparison was unfeasible until group catalogs were built with overdensities roughly matching the ``virial"  criterion for dynamical equilibrium (overdensities relative to the critical density, $3H^2/(8\pi\,G)$, on the order of 100 in the low-redshift Universe). An example of this type of catalog is the popular \cite{Yang+07} group catalog.
Using a SAM,
\cite{Zandivarez+14} 
found no difference in the size and shape of CGs compared to those of regular groups with similar projected separation between the first- and second-ranked galaxies.
On the other hand,
\cite{Zandivarez+22} have claimed a slight but significant deficiency of faint ($M_r > -17$) galaxies in CGs compared to regular groups. 

The properties of a CG also depend on its position relative to its host group.
\cite{DiazGimenez+15} found that CGs embedded in regular groups have smaller sizes and a higher surface brightness than non-embedded CGs.
More recently,
\cite{Zheng&Shen21} found that roughly half of their CGs are embedded in regular groups, whereas roughly one quarter are ``Isolated" compact groups, leaving almost another quarter of their compact groups 
``Split" between several regular groups. 
They also found that over half of the embedded CGs dominate the luminosity of the parent group. They call these ``Predominant," while those not dominating the group luminosity are called ``Embedded" and have higher velocity dispersions than both the Isolated CGs and non-compact groups (which have similar velocity dispersions as the Isolated CGs).\footnote{The capitalized words are defined in section \ref{sec:locvsZheng}}

This article aims to go beyond the pioneering analysis of \cite{Zheng&Shen21} by 
addressing the following points:
1) determining how the relative populations of classes of CGs defined by \cite{Zheng&Shen21} differ when 
only considering compact groups of at least four galaxies (of concordant luminosities), to better conform with the original intent of \cite{Hickson82};
2) finding out where the embedded CGs (i.e., in both Predominant and Embedded classes) are located relative to their parent groups; 
3) studying how the CG properties and their correlations compare to those of previous catalogs.

To answer these questions, we adopted the CG catalog of \cite{DiazGimenez+18}, which as mentioned above is the largest fully complete sample of CGs in magnitude range and available spectroscopy. More precisely, we considered the majority of the CGs that have exactly four member galaxies. 
We compare the compact groups to control samples formed from the regular group catalog of \citet{Lim+17}.
When comparing CG and regular group samples, one should note
that \citet{Lim+17} and \citet{DiazGimenez+18} did not adopt 
the same algorithms to build their groups, nor did they start from the same set of galaxies.
In fact, some HMCGs are split among several (up to four) Lim groups.\footnote{Three HMCGs are split between four Lim groups: HMCG~359 ($N= 6$ members) has three galaxies in $N=4$ Lim group 10218 and three galaxies in three separate Lim groups of a single galaxy; HMCG~415 ($N=5$) has two galaxies in a Lim group of two galaxies and three galaxies in three separate Lim groups of a single galaxy; as another example, the four galaxies of HMCG~445 are split into four Lim groups with $N=11, 8, 3$, and 1, respectively.}

We present the CG data and the control samples in Sect.~\ref{sec:data}, discuss the locations of CGs relative to regular groups in Sect.~\ref{sec:environment}, and describe the CG structure and fundamental properties in Sect.~\ref{sec:distrib}.
In Sect.~\ref{sec:concdisc}, we present and discuss our conclusions.

Throughout this article, following \cite{Tempel+17}, we adopt cosmological parameters for a flat Universe with $\Omega_\mathrm{m}=0.308$, $N_\mathrm{eff}=3.15$, and $h=0.678$ 
\citep{Planck_collaboration+15_cosmopars}, and all our logarithms are base 10.

\section{Data}	
\label{sec:data}

\subsection{Galaxies}
\label{sec:Tempel}
We obtained galaxy properties from the Friends-of-Friends group catalog that \cite{Tempel+17} extracted from Data Release 12 of the SDSS Main Galaxy Sample. \citeauthor{Tempel+17} made a great effort to remove poorly measured galaxies. Their catalog\footnote{\url{http://cdsarc.u-strasbg.fr/viz-bin/qcat?J/A+A/602/A100}}
contains 584\,449 galaxies up to a redshift of $0.2$ in the frame of the cosmic microwave background (CMB).

Absolute magnitudes are derived from apparent magnitudes using distances in the Planck cosmology \citep{Planck_collaboration+15_cosmopars}, Galactic extinction from \cite*{Schlegel+98}, k-corrections from \cite{Blanton&Roweis07}, and evolutionary corrections from \cite{Blanton+03}. 
Although k+e corrections in \cite{Tempel+17} are not described, it is likely that they follow those in \cite{Tempel+14}, which are based on \citet{Blanton&Roweis07} for k and \citet{Blanton+03} for e, assuming a distance-independent luminosity function. 
We derived $r$-band luminosities assuming a solar absolute magnitude of $\mathrm{M}_{\odot,r}=4.68$.\footnote{From \url{https://www.sdss4.org/dr12/algorithms/ugrizvegasun}.}
All matching of SDSS-based catalogs shown in Table~\ref{tab:sample_summary} (see Sect.~\ref{sec:control}) are based on SDSS galaxy \texttt{OBJID}.

\subsection{Regular groups}\label{catalogs}

We use the group catalog\footnote{Available at the ``SDSS DR7" section at \url{https://gax.sjtu.edu.cn/data/Group.html}} identified by \citet{Lim+17} from SDSS DR7,  who essentially followed the group-finding algorithm of \cite{Yang+07}. 
The algorithm of \citeauthor{Lim+17} works in the five following steps: 1) it starts by assigning a single-galaxy group to each galaxy; 2) it determines the virial radius and velocity dispersion of each group using the group luminosity - total mass relation determined from a cosmological hydrodynamical simulation; 3) it estimates a ``factor" (related to the probability of membership) linking each galaxy to  each group; 4) it assigns each galaxy to the group with its highest factor; 5) it iterates from step 2, modified to use abundance matching between group luminosity and the halo mass function measured form cosmological simulations for groups of more than one galaxy.
We work with their sample \texttt{SDSS(L)group.dat} that comprises $446\,495$ groups with one or more members containing $586\,025$ galaxies. The ``\texttt{L}" version means it is constructed with galaxies that have spectroscopic redshifts, using Proxy-L (based on galaxy luminosities) to estimate halo masses. In this sample, $388\,819$ groups have a single galaxy, while $12\,213$ have four or more members. 

We first removed the twelve duplicate galaxies from the catalog. We then found that $9420$ galaxies in these groups do not belong to the \cite{Tempel+17} galaxy catalog. Since we are using galaxy properties from \citeauthor{Tempel+17}, we removed from the group catalog $8087$ groups that host those galaxies. We are then left with  $438\,408$ groups containing $573\,572$ galaxies. Hereafter, this restricted sample will be named as \LT{} catalog.\footnote{
We do not use the grouping of \citeauthor{Tempel+17}, because it is based on Friends-of-Friends, and appears considerably less reliable than the \cite{Yang+07} algorithm (as shown by \citealt{Duarte&Mamon15}).}

\subsection{Compact groups}
\label{sec:CGs}
\subsubsection{Sample}
\label{sec:CGsample}
We use the sample of Hickson Modified Compact Groups (HMCG) identified by \cite{DiazGimenez+18} on the \cite{Tempel+17} galaxy catalog.  The HMCG sample contains 462 compact groups, among which 406 are not flagged as dubious by the authors.  By construction, not all the galaxies in HMCG lie in the \citeauthor{Tempel+17} galaxy catalog, because \cite{DiazGimenez+18} had added 63 additional galaxies, with redshifts from the NASA Extragalactic Database (see their table~B2).

We selected the 284 compact groups among the 406 reliable HMCGs that contain exactly four galaxies.
Among these 284 groups, we discarded all those containing at least one galaxy not present in the \LT{} sample, 
leaving us with 226 compact groups of four galaxies belonging to the \LT{} sample.
The reasons for such a high fraction of discarded groups ($62/\,284=22\%$) are the 63 galaxies added by \cite{DiazGimenez+18} and the 7000 galaxies added by \cite{Tempel+17}  that are not in the SDSS Main Galaxy Sample
and the  $600\,458-586\,025=14\,433$ galaxies that are not part of the group catalog of Lim et al. 2017 (field galaxies). Rest-frame absolute magnitudes, CMB redshifts and angular coordinates for galaxies in this subsample are extracted directly from \cite{Tempel+17}.

\begin{figure}
\centering
 \includegraphics[width=\columnwidth]{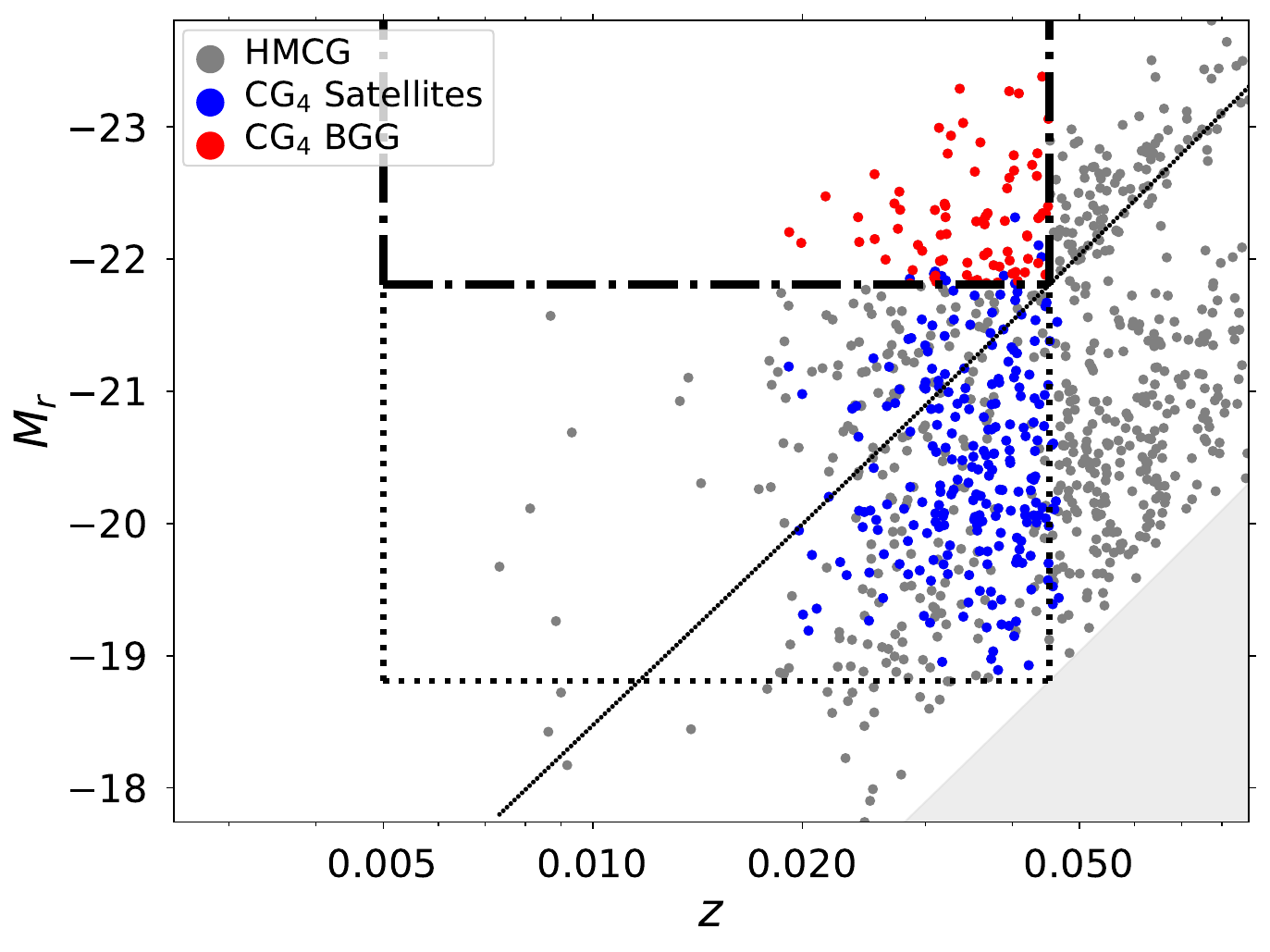}
 \caption{Absolute $r$-band magnitude versus redshift for galaxies in the HMCG compact group sample and in the \CG{} subsample.
 The colors are gray for all HMCG galaxies, red for \CG{} BGGs and blue for \CG{} satellites. The shaded gray region 
represents galaxies fainter than the SDSS flux limit ($r> 17.77$). The dotted line denotes our doubly complete HMCG sample, while the thick dash-dotted lines shows the magnitude and redshift limits for the BGGs.  The oblique line is three magnitudes above the SDSS flux limit and is our limit for BGGs in the first-step \CG{} selection process.
}
 \label{fig:CG_Selection}
\end{figure}

We choose, here, to compare the properties of compact groups and regular groups, controlling for multiplicity (i.e., richness) and galaxy luminosity. 
We build a subsample of 4-member CGs  that is doubly complete in volume and luminosity to prevent selection effects, in particular on the magnitude range, since the Hickson CG multiplicity criterion imposes a minimum of four galaxies within three magnitudes from the BGG, i.e. $r_\mathrm{BGG} < 14.77$ for our SDSS-based sample. 
We first required $z_\mathrm{BGG} > 0.005$ to avoid galaxies whose unknown peculiar velocities will contribute sufficiently to the redshift that the distance inferred from the redshift is too uncertain, leading to too uncertain luminosities.
The number of groups in such doubly complete subsamples depends on the choice of maximum redshift or equivalently on minimum luminosity.
Starting from the 226 CGs of four galaxies, we recover a maximum of 78 CGs (therefore containing 312 galaxies) for $M_{r,\mathrm{BGG}} < -21.81$, corresponding to $z_\mathrm{BGG} < 0.0452$ for $r_\mathrm{BGG}<14.77$.
 This selection is illustrated in Figure~\ref{fig:CG_Selection}.
We call \CG{} this sample of 78 compact groups of four galaxies.

\subsubsection{Comparison to other compact group samples}
\label{sec:compareCGs}

\begin{figure}
\centering
 \includegraphics[width=\hsize]{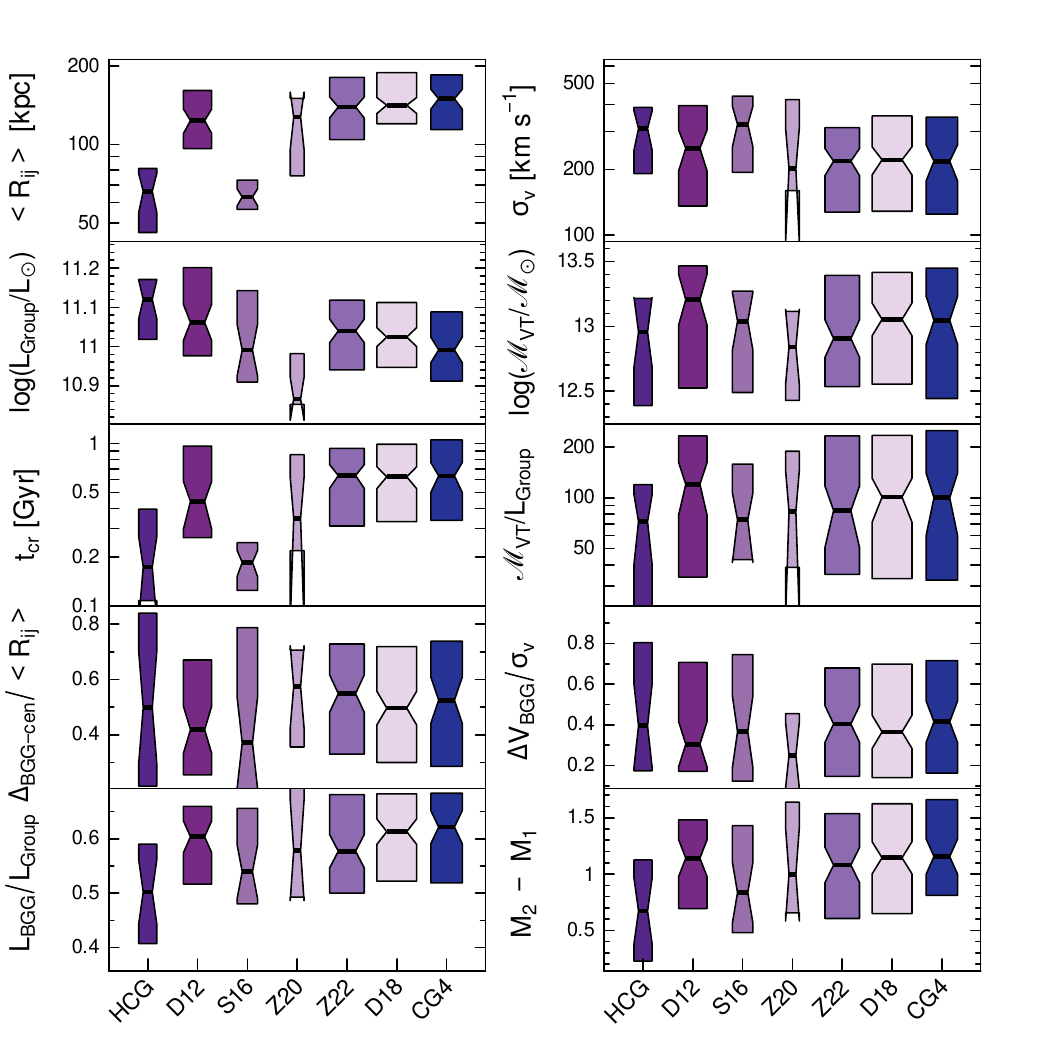}
 \caption{Comparison of \CG{} properties with samples of compact quartets built from other published compact group catalogs, with the same selection in redshifts and magnitudes: $0.005 \leq z_\mathrm{BGG} \leq 0.0452$, $M_\mathrm{BGG}\leq-21.81$, and $M_4-M_1\leq 3$.
 From top left to bottom right, the properties are median of the projected inter-galaxy separations, radial velocity dispersion, 
 total group luminosity, 
 group virial theorem mass, 
 crossing time,
 mass-to-light ratio, 
 BGG relative offset (normalized distance of the BGG to the centroid of the group), 
 BGG relative velocity offset (normalized distance in radial velocity of the BGG to the group radial velocity dispersion, top-right panel), 
 BGG luminosity fraction, 
 and 
 absolute magnitude difference between the two brightest galaxy members.
The abscissa indicate the compact group catalogs:
 HCG: \protect\cite{Hickson+92},
 D12: \protect\cite{DiazGimenez+12},
 S16: \protect\cite{Sohn+16},
 Z20: \protect\cite{Zheng&Shen20},
 Z22: \protect\cite{Zandivarez+22},
 D18: \protect\cite{DiazGimenez+18}, and
 CG4: this work.
 \label{fig:OtherSamples}
 }
\end{figure}

 We now compare our \CG{} catalog to previously published compact group catalogs: 
 \cite{Hickson+92} (HCG)\footnote{\url{https://vizier.cds.unistra.fr/viz-bin/VizieR?-source=VII/213}},
\cite{McConnachie+09} (M09)\footnote{\url{https://vizier.cds.unistra.fr/viz-bin/VizieR?-source=J/MNRAS/395/255}}, \cite{DiazGimenez+12} (D12)\footnote{\url{https://vizier.cds.unistra.fr/viz-bin/VizieR?-source=J/MNRAS/426/296}}, \cite{Sohn+15} (S15)\footnote{\url{http://astro.snu.ac.kr/~jbsohn/compactgroups/}}, \cite{Sohn+16} (S16)\footnote{\url{https://vizier.cds.unistra.fr/viz-bin/VizieR?-source=J/ApJS/225/23}}, \cite{Zheng&Shen20} (Z20)\footnote{\url{https://vizier.cds.unistra.fr/viz-bin/VizieR?-source=J/ApJS/246/12}},
\cite{Zandivarez+22} (Z22)\footnote{\url{http://vizier.cds.unistra.fr/viz-bin/VizieR-2?-source=+J/MNRAS/514/1231}},
and \cite{DiazGimenez+18} (D18)\footnote{\url{https://vizier.cds.unistra.fr/viz-bin/VizieR?-source=J/A+A/618/A157}. 
}. 

For a fair comparison, we filtered these catalogs to groups of exactly four redshift-concordant galaxies with our limits in distance ($0.005 \leq z_\mathrm{BGG} \leq 0.0452$) and luminosity ($M_\mathrm{BGG} \leq -21.81$), as well as a magnitude concordance criterion ($M_4-M_1\leq 3$). This led  to zero M09 groups and only one S15 group, so we discarded both M09 and S15 samples. This left us with 24 compact quartets in HCG, 
59 in D12, 31 in S16, 14 in Z20, 88 in Z22, 117 in D18, and 73 in 
CG4.\footnote{In this section, the three-magnitude maximum allowed for the difference between BGGs and satellites in  \CG{} groups concerns absolute magnitudes, instead of apparent magnitudes as used in other sections. There are five \CG{}s with maximum absolute magnitudes gaps (3.02, 3.06, 3.07, 3.18 and 3.25).}

We calculated the following group properties for all samples in the same way:
median projected galaxy separation $\langle R_{ij}\rangle$,
group velocity dispersion $\sigma_v$,
group luminosity $L_\mathrm{group}$,
group virial theorem mass ${\cal M}_\mathrm{VT}$,
relative BGG offset $\Delta_\mathrm{BGG-cen}/\langle R_{ij}\rangle$,
relative BGG velocity offset $\Delta v_\mathrm{BGG}/\sigma_v$,
relative BGG luminosity $L_\mathrm{BGG}/L_\mathrm{group}$,
magnitude gap $M_2-M_1$,
crossing time $t_\mathrm{cr}$, 
group virial theorem mass to $r$-band luminosity ratio ${\cal M}_\mathrm{VT}/L_r$.

For the catalogs that only provided observer-frame apparent magnitudes, we used the k-corrections recommended by \cite{Chilingarian+12} to obtain rest-frame absolute magnitudes at $z=0$ and $\msun =4.68$. For CG4 we use the rest-frame absolute magnitudes of the galaxy members, $M_r$, provided by \cite{Tempel+17}.

 For each group, 
we converted the angular separations, $\theta_{i,j}$, into distances projected on the plane of sky, $R_{i,j}$ using 
\begin{equation}
    R_{i,j} = \theta_{i,j}\,D_\mathrm{A}(z_\mathrm{group}) \ ,
    \label{Rijofz}
\end{equation}
where $D_\mathrm{A}(z)$ is the cosmological angular separation
distance.\footnote{We neglected peculiar velocities, as we can expect
the BGG velocity to be close to the velocity of the group's center of mass.} 
 We computed the line-of-sight velocities of galaxies relative to their group using
 \begin{equation}
 v_i = c\,{\left(z_i - z_\mathrm{group}\right)
     \over 1+z_\mathrm{group}}
    \ .
     \label{dv} 
 \end{equation}
 The line-of-sight velocity dispersion, $\sigma_v$,
 was then estimated using the gapper algorithm \citep{Wainer1976}, which is the most efficient for measuring radial velocity dispersion for small samples \citep*{Beers+90}. 
 Finally, from the rest-frame absolute magnitudes of the galaxy members, 
 we computed the group luminosities from the sum of luminosities of individual galaxies, $L_{r,\mathrm{group}} = \sum \mathrm{dex}[-0.4\,(M_r\!-\!\msun)]$.
Furthermore, the virial theorem mass is
(\citealt*{HTB85}, written as in \citealt{DiazGimenez+12})
\begin{equation}
{\cal M}_\mathrm{VT} = 3\pi {R_\mathrm{h}\,\sigma_v^2\over G} =
  2.192\times 10^{6} \,
  \left(\frac{ R_\mathrm{h}}{\mathrm{kpc}}\right)
  \,\left( \frac{\sigma_v}{\mathrm{km\,s^{-1}}}\right)^2 
  \,\msun \ ,
  \label{M_VT}
\end{equation}
where $R_\mathrm{h}=\left\langle R_{ij}^{-1}\right\rangle^{-1}$ is the harmonic mean projected separation.
Following \cite{DiazGimenez+12}, the crossing time is defined as
\begin{equation}
t_\mathrm{cr} = {\left\langle r_{ij}\right\rangle\over \sigma_{v,\mathrm{3D}}} = 
{\pi\over 2\,\sqrt{3}}\,
{\left\langle R_{ij}\right\rangle\over \sigma_v} = 0.887\,
\left\langle {R_{ij}\over \mathrm{kpc}}\right\rangle \,\left(\frac{\mathrm{km\,s^{-1} }}{\sigma_v} \right) \mathrm{Gyr} \ ,
\label{tcr}
\end{equation}
where $r_{ij}$ are the 3D separations. 

The properties of the group samples are compared in the boxplot diagrams
of Fig.~\ref{fig:OtherSamples}.\footnote{In the boxplot diagrams, the top and bottom lines of the boxes represent the 25th and 75th percentiles of the distributions, while the wrists of the boxes represent the medians. 
Notches display the confidence interval (95\% confidence level)
symmetrically around the medians. When comparing distributions, if the
notches of two boxes do not vertically overlap, there is a statistically significant difference between the medians (\citealt*{boxplot78}; \citealt{boxplot14}).
For skewed distributions or small-sized samples, it might happen that the CI is wider than the 25th or 75th percentile. Therefore, the plot will display an ``inside out" shape. 
The lines extending from the boxes are called whiskers (not shown in Fig.~\ref{fig:OtherSamples} but in Fig.~\ref{fig:Distributions_Groups}). 
The boundary of the whiskers is based on the 1.5 inter-quartile range (IQR) value. The whiskers extend from the bottom (resp. top) of the boxes to the lowest (resp. highest) data point that falls within 1.5 times the IQR.
The whisker lengths might not be symmetrical since they must end at an observed data point.}
By construction, the properties of \CG{} resemble those of D18, from which it was built (but in the smaller zone covered by  \citealt{Lim+17}). However, the different catalogs were built using different criteria for group compactness and possible isolation.
Apart from the samples of \cite{Hickson+92} (HCG) and \cite{Sohn+16} (S16), whose compact quartets are much smaller (their median separations are half those of the other samples), the other samples of compact quartets are in fair agreement with our \CG{} sample. 
Also, the HCG sample shows the smallest magnitude gap and lowest dominance of the BGG. Previous works \citep{Prandoni+94,DiazGimenez&Mamon10} had noted that Hickson's visual selection was biased to avoid groups whose magnitude gap is close to the limit of the selection.
One other standout sample is that of \cite{Zheng&Shen20}, whose quartets have lower total luminosities and marginally lower relative BGG velocity offsets than those of the other samples. 

\subsection{Control samples of regular groups}\label{control}
\label{sec:control}
We built three control samples of regular groups. We began by selecting a subsample of \LT{} groups with the following steps: 
\begin{enumerate}
    \item We discarded all \LT{} groups having their BGG outside of the redshift limits of our \CG{} selection ($0.005 < z_\mathrm{BGG} < 0.0452$) to ensure the same level of completeness as for the \CG{} groups. This left us with 46\,279 groups containing 68\,834 galaxies. 
    \item In this redshift range, we selected groups in the same $M_r$ limits as of our \CG{} selection. This yielded 1950 groups (hereafter LT${}_{z,Mr}$), containing 14\,265 galaxies, having their BGG brighter than $-21.81$. \label{LtzMr}
    \item We discarded groups with fewer than four galaxies, which led to 1042 groups containing 12\,516 galaxies. \label{LT4}
    \item We retained those groups with at least three satellites whose luminosities lie within 3 absolute magnitudes from their BGG, leaving 765 groups (73\% of our previous sample), containing 11\,087 galaxies (i.e., an average of 14 galaxies per group). This selection is hereafter named Parent Control catalog (PC). Among 765 PC groups, 68 (9\%) 
     contain at least one galaxy from \CG{}: 
  57 PC groups have four galaxies from \CG, 5 have three galaxies, 3 have eight, one has six galaxies, one has two and one has a single galaxy. 
     We will discuss in Sect.~\ref{sec:environment} how \CG{}s are located within their parent groups.\label{pcgroup}
 \end{enumerate}

\begin{figure}
\centering
 \includegraphics[width=\columnwidth]{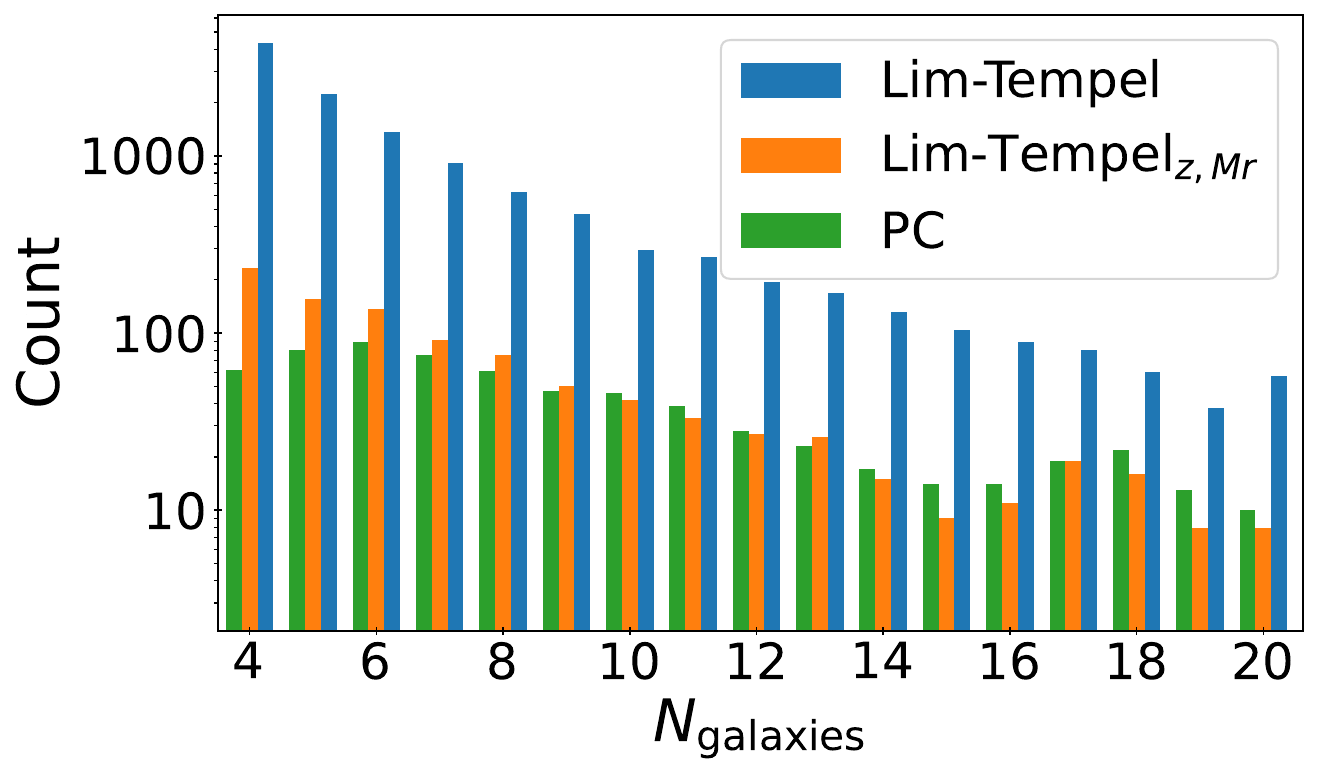}
 \caption{Multiplicity functions of the parent group samples: LT, LT$_{z,Mr}$ (step \ref{LtzMr} in Sect.~\ref{control}), and PC (step \ref{pcgroup} in Sect.~\ref{control}) in the range $4 \leq N_\mathrm{galaxies} \leq 20$
 ($N_\mathrm{galaxies}$ reaches 538 for PC).
The 62
 PC groups of four correspond to the \PCF\ sample. }
 \label{fig:PC_richness}
\end{figure}

Figure ~\ref{fig:PC_richness} compares the multiplicity functions of the LT$_{z,Mr}$ (step \ref{LtzMr} in Sect.~\ref{control}) and PC (step \ref{pcgroup} in Sect.~\ref{control}) groups, in the range $4 \leq N_\mathrm{galaxies} \leq 20$. 
\LTzMr{} contains roughly one order of magnitude fewer galaxies than \LT{} for multiplicities comprised between 4 and 15. The multiplicity function of \PCF{} groups is also roughly ten times lower than that of LT groups, and it is even more than ten times lower for groups of fewer than six members.

From PC groups, we also built three control samples of regular groups of four galaxies: 
\begin{itemize}
    \item Regular groups of four galaxies (\PCF{}), which  comprises 62 groups. 
    \item \CB{} comprise the 765 PCs in which only the BGG and the three other brightest galaxies are selected (the subscript B stands for ``brightest").

    \item \CC{} is composed of the 765 PCs in which only the BGG and the three closest galaxies  to the BGG (in projection) are selected (the subscript C stands for ``closest").
\end{itemize}

\begin{table*}
 \caption{Summary of the group samples used in this study. 
 }
    \centering
    \parindent=0pt
    \tabcolsep=3pt
    \begin{tabular}{p{2cm} p{3cm} p{12.5cm}}
    \hline
    \hline
        \head{Name} & \head{Number of groups} & \head{Description} \\
    \hline
         \ \CG&\ \ \ \ \ \ \ \ \ \, 78 & Compact groups with exactly four members 
              in a volume-limited catalog extracted from \cite{DiazGimenez+18}\\
   \rule{0pt}{3ex}     
        \LT & \ \ \ \ \ \,438\,408 
        \newline 
        (573\,572 galaxies)& Intersection of the \citet{Lim+17} and \cite{Tempel+17} group catalogs\\ 
\rule{0pt}{3ex}  
\LTzMr & \ \ \ \ \ \ \,1950 \newline (14\,265 galaxies)& \LT{} groups  in our volume- and luminosity-limited subsample \\
\rule{0pt}{3ex} 
        PC 
        & \qquad\ \ 765\newline (11\,087 galaxies)
        & \LTzMr{} groups of at least four galaxies within three magnitudes of the brightest\\
\rule{0pt}{3ex}     
         \PCF &\ \ \ \ \ \ \ \ \ \,56 & PC groups of exactly four galaxies, excluding those whose galaxies are identical to those of a \CG{} \\
\rule{0pt}{3ex}  
         \CB &\ \ \ \ \ \ \ \  699 & Control sample from PC with the four brightest galaxies in each group, excluding those with galaxies in common with \CG{}s\\
\rule{0pt}{3ex}     
         \CC &\ \ \ \ \ \ \ \ 704 & Control sample from PC with the brightest and its three closest (in projection) galaxies, i.e. the core of the PC group, excluding those with galaxies in common with \CG{}s\\
    \hline
    \end{tabular}
   
    \label{tab:sample_summary}
\end{table*}

Finally, we excluded the six \PCF{}s whose galaxies are identical to
those of a \CG. Thus, six out of 78 \CG{}s (8\%) are a Lim group, and among 62 potential \PCF{}s, 6 (10\%) are compact.  We also excluded the control groups containing at least one galaxy belonging to a \CG{} (66 in \CB, 61 in \CC).  Our final control samples have 699 groups in \CB{}, 704 in \CC{} and 56 in \PCF{}.
The five group samples used in this study are summarized in Table~\ref{tab:sample_summary}.

\begin{figure}
\centering
 \includegraphics[width=\columnwidth]{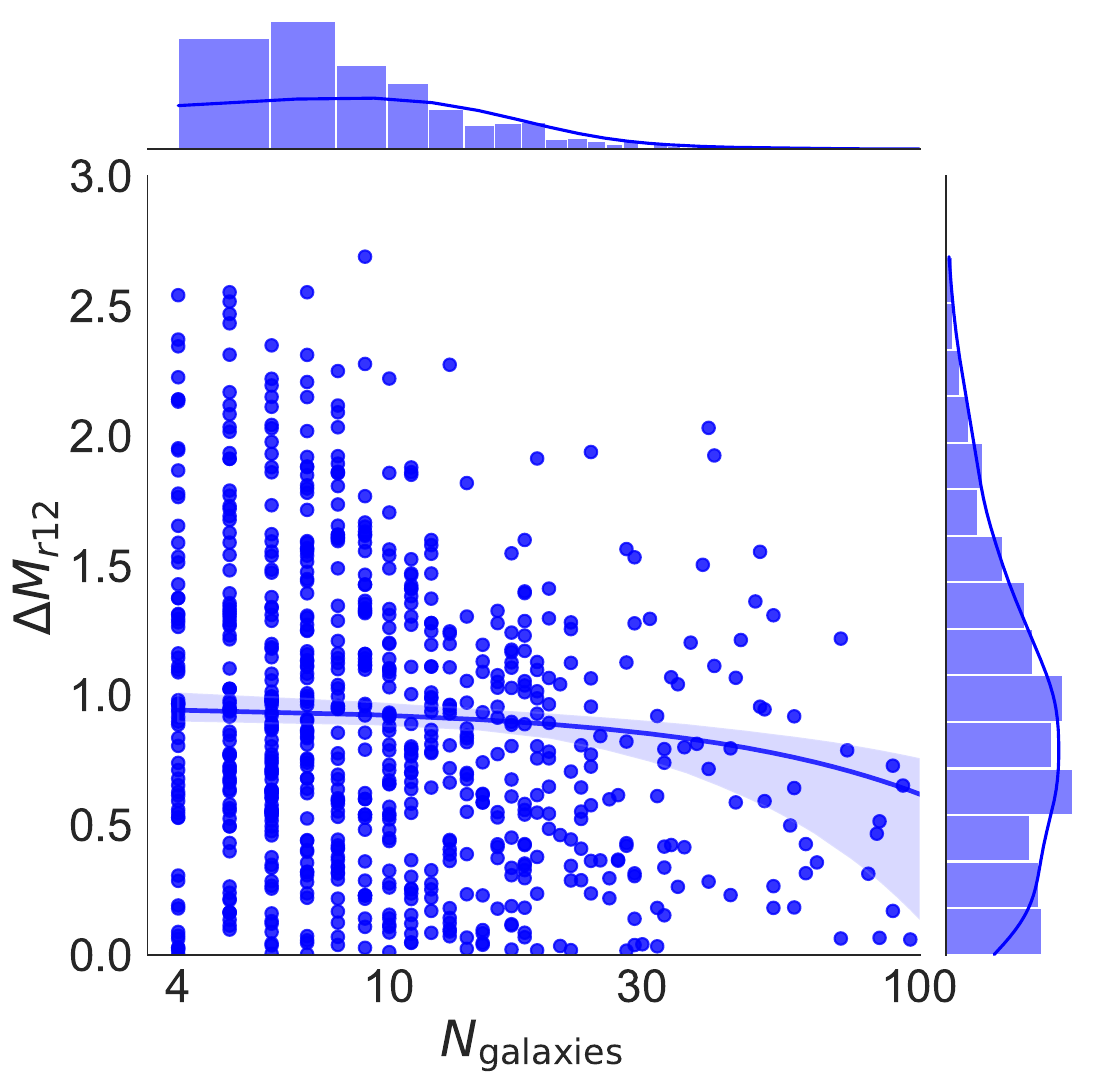}
 \caption{Absolute magnitude difference between the BGG and the second brightest galaxy of each group versus group multiplicity for PC groups. 
 }
 \label{fig:GapMag_vs_Ngal}
\end{figure}

The goal of creating control group samples having the same number of galaxies as \CG{}s is to avoid common multiplicity effects. For example, one can expect that the more galaxies in a group, the more likely there will exist a galaxy having a magnitude closer to the BGG magnitude in this group. Fig.~\ref{fig:GapMag_vs_Ngal} shows the relation between absolute magnitude gap 
and multiplicity for the PC groups.
The visual impression is that the gap decreases only weakly with increasing multiplicity. But a Spearman rank  test finds a correlation coefficient of --0.2, with a probability of occurring by chance of less than $10^{-7}$. Hence, our need to control for multiplicity.

\label{sec:Split}

\section{Location of compact groups within their parent group}
\label{sec:environment}

\subsection{\CG{}s within groups from the \LT{} catalog and its doubly complete subsample}
Since, by construction, the \CG{} sample is restricted to compact groups whose members are all members of the \LT{} sample, we can analyze the associations between the two group samples.

\LT{} groups hosting \CG{} galaxies have a variety of multiplicities, from a single galaxy to 538 galaxies, with a median multiplicity of 10 galaxies.
On the other hand, out of 78 \CG{} groups, 62 are fully embedded in a \LT{} group, 14 \CG{} groups are split into two \LT{} groups, and two \CG{} groups are split over three \LT{} groups. Having compact groups split over several ordinary groups can only happen because, as stated in Sect.~\ref{sec:Split}, the galaxy group finders of \citet{Lim+17} and \citet{DiazGimenez+18} did not use the same prescriptions.
For instance, 11 \LT{} single-galaxy groups  are actually made of a \CG{} galaxy). We will explore this shredding
in Sect.~\ref{sec:locvsZheng}.

Four of the \LT{} groups contain two \CG{} BGGs, while the \LT{} hosts of the remaining $78-4\times2 = 70$ \CG{}s contain a single \CG{} BGG.
However, only 68 of the 78 \CG{} BGGs are the most luminous galaxy (i.e., BGG) of their host \LT{} group.

Among the 1950 LT$_{z,Mr}$ groups, 77 (4\%) contain at least one galaxy of a \CG: 
two have a single galaxy from a \CG,
six have two galaxies, 
eight have three, 
57 have four, 
one has six, 
and
three have eight.

\subsection{\CG{}s within PC groups\label{sec:CG_in_PC}}

Out of 78 \CG{}s, 72 have at least one galaxy within a PC group: 63 have all their galaxies embedded in a PC group (62 within a single group, and one split between two groups), six have only three galaxies belonging to a PC group, and three have only two galaxies in a PC. Therefore, $4\times (78\!-\!72) + 1\!\times\!6 + 3\!\times\!2 = 36$ \CG{} galaxies are missing from PCs: $4 \times(78-72)=24$ from \CG{}s with no matching galaxies with PCs and 12 within \CG{}s that have at least one galaxy in a PC. The $78-72=6$ \CG{}s having no galaxy in a PC group are split between two or more \LT\ groups containing fewer than four members.
The 72 \CG{}s having a galaxy in a PC group belong to 68 such PC groups. Among our 78 \CG{}s, 70 of their BGGs belong to a PC group.
Among these 70, 60 are the BGGs of their host groups.

\label{sec:CGBGG_in_PC}

For each PC group, we computed its (projected) centroid location\footnote{We define centroid by converting the equatorial coordinates to a cartesian frame on the unit sphere, taking the means and converting back to the equatorial frame.} and a proxy for its virial radius. The \citeauthor{Lim+17} catalog provides group masses ${\cal M}_{180}$ defined at the radius of the sphere whose mean density is 180 times the mean density of the Universe at the group redshift. We converted it into the more usual ${\cal M}_{200\mathrm{c}}$ and $R_{200\mathrm {c}}$, where $R_{200\mathrm{c}}$ is the virial radius where the mean density is 200 times the critical density of the Universe, and ${\cal M}_{200\mathrm{c}}$ the corresponding mass. For this, we used the procedure in Appendix~\ref{sec:conversion}. 

\begin{figure}
 \includegraphics[width=\columnwidth]{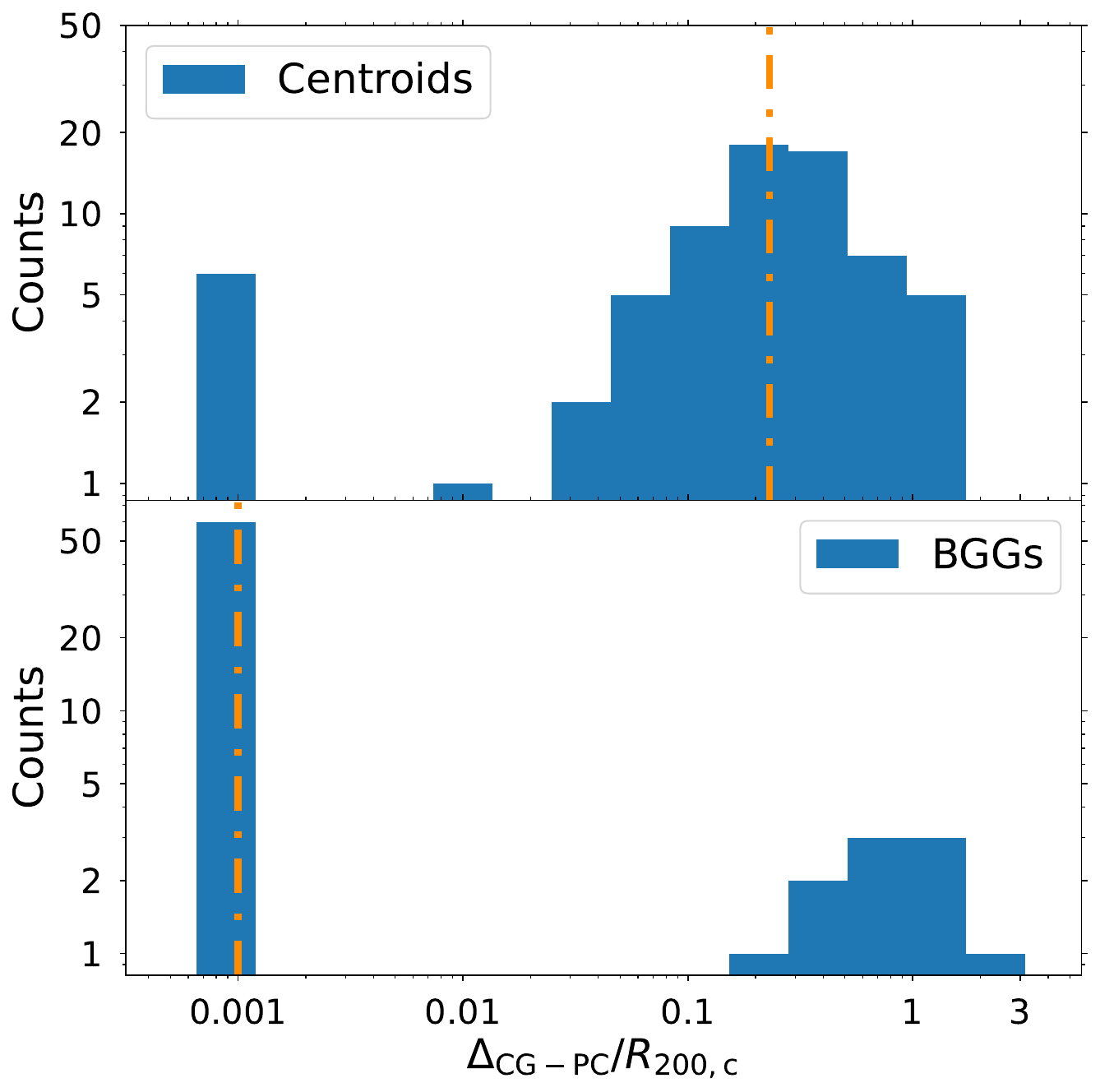}
 \caption{Distributions of offsets of the \CG{} positions relative to their host PC group, computed between group centroids ({\textbf{top}}) and between brightest group galaxies ({\textbf{bottom}}) in virial units.
 The vertical lines indicate the medians (0.23 for the centroids and 0 for the BGGs).
In both panels the null values were clipped to 0.001. The peaks at 0.001 in the upper panel correspond to perfect matches between the \CG\ and \PCF\ groups, while those in the bottom panel correspond to the \CG\ and PC groups sharing the same BGG. 
}

 \label{fig:Virial_offset}
\end{figure}
Fig.~\ref{fig:Virial_offset} displays the separations between the centroids of \CG{}s and their host PC groups (top panel), and the separations between \CG{} BGGs and the PC host group BGGs (bottom panel), both expressed in units of the host group virial radii. 
Three PC groups contain two BGGs of different \CG{}s. Those three cases are therefore counted twice.
The upper panel of Fig.~\ref{fig:Virial_offset} indicates the CG centroid is typically offset by 0.23 PC virial radii from the PC centroid.
This preference for CGs to be located deep in their host groups was previously noted by \cite{Taverna+23} using very different methods.
The lower panel indicates that 60 \CG\ BGGs out of 70 (86\%) are coincident with the PC BGG (as previously mentioned), while six are the second brightest, one is the third brightest, two are the fourth brightest, and one is the fifth brightest.
This shows that \CG{} BGGs tend to lie, in projection, in PC group cores.
On the other hand, five out of the 70 \CG{} BGGs lie beyond the PC virial radius.

\subsection{Comparison to Zheng and Shen}
\label{sec:locvsZheng}
\cite{Zheng&Shen21} performed one-way membership matching of the CGs of their recent catalog \citep{Zheng&Shen20} relative to the \cite{Yang+07} groups in order to analyze the location of the compact groups relative to their parent groups and to split their CGs into four classes.
We applied their classification to the \CG{}s using their nomenclature.

For our selection, all \CG{} galaxies belong to \LT{} groups (we note that 18 \CG{}s contain galaxies belonging to \LT\ groups with fewer than four galaxies). Among our 78 \CG{} groups, we note the following characteristics:
\begin{itemize}
    \item Sixteen (21\%) are ``Split" between several \LT{} groups, similar to the 21\% of Split CGs found by \cite{Zheng&Shen21}. Among these 16 groups, 14 are split in two parent groups (four \CG{}s in a 2+2 configuration and ten are in a 3+1 configuration), while two are split between three parent groups (in a 2+1+1 configuration). Among the 32 \LT{} groups hosting Split \CG{} fragments, one group hosts Split \CG{} galaxies coming from two different \CG{}s and one hosts galaxies coming from three different Split \CG{}s.\footnote{There are only 32 \LT{} groups hosting Split \CG{} fragments instead of the expected $14 \times 2 + 2\times 3=34$ since one LT group hosts fragments from three different \CG s.}
    \item Six (8\%) are ``Isolated", i.e. they form a whole \LT{} group by themselves, while \citeauthor{Zheng&Shen21} found 27\% of Isolated CGs.
    \item Nineteen (24\%) are ``Predominant", i.e non-Isolated, non-Split, and accounting for half or more of their host group total $r$-band luminosity, while \citeauthor{Zheng&Shen21} reported 26\%.
    \item  Thirty-seven (47\%) are ``Embedded", i.e non-Isolated, non-Split, and accounting for less than half their host group total $r$-band luminosity, which is much higher that the 23\% found by \citeauthor{Zheng&Shen21}.
\end{itemize}
Barnard tests between one class and all the others indicate that the lower fraction of Isolated groups and higher fraction of Embedded groups  among \CG{}s relative to \citeauthor{Zheng&Shen21} groups are both highly significant ($p<10^{-4}$ and $p<10^{-5}$, respectively).
Therefore, while \cite{Zheng&Shen21} find roughly a quarter of each of their groups belonging to each category, our sample shows, in comparison, an excess of Embedded groups and a lack of Isolated groups.

The higher median multiplicity of the \CG{}s (4) compared to the \citeauthor{Zheng&Shen20} groups (3) explains these two differences.
All group catalogs show a rapidly decreasing multiplicity function. This can be explained by the expected correlation between group multiplicity and mass on one hand and the declining halo mass function (predicted first by \citealt{Press&Schechter74}, and confirmed later by numerous studies based on cosmological simulations). Triplets are considerably more frequent than quartets (2.4 times in \LT\ and 7.6 times in \citealt{Zheng&Shen20}). By extension, Isolated triplets should be much more frequent than Isolated quartets.
In fact, 
higher-multiplicity groups must be rarer than lower-multiplicity ones. Assuming Poisson statistics for the group multiplicity, i.e. the number of groups of multiplicity $N$ is $P_N = \exp(-\left\langle N\right\rangle)\,\left\langle N\right\rangle^N/N!$, the ratio of counts of groups of $N$ members over counts of groups of $N$+1 members is $(N+1)/\langle N\rangle$. Thus, the ratio of numbers of groups of $N=3$ members to groups of four is expected to be $4/\left\langle N\right\rangle$,
i.e. $\simeq3.1$ times more groups of three members than of four members for catalogs with $\langle N\rangle = 1.3$, such as \LT\ (see Table~\ref{tab:sample_summary}). 
This predicted ratio of 3.1 is consistent with the \citeauthor{Zheng&Shen20} (27\%) over \CG\ (8$\pm$2\%) fractions of Isolated compact groups (where the uncertainty is from Poisson statistics).

The relative preference for Embedded \CG{}s compared to those of \citeauthor{Zheng&Shen20} may be caused by the lack of dwarf galaxies in the typically higher redshift \citeauthor{Zheng&Shen20} CGs compared to our \CG{}s, selected at the same flux limit.

\section{Properties}
\label{sec:distrib} 

We now compare the properties of the different group samples and subsamples. We begin with the global group properties, and continue with the brightest group galaxy (BGG) location within their group in position, velocity and luminosity space. We then discuss group properties according to their location within their host groups and study correlations in \CG{}s.
In a forthcoming article, we will compare and analyze the galaxy populations in the different samples.

\subsection{Group properties}
\label{sec:properties}

\begin{table*}  
\caption{Comparison of median properties between compact group quartets and control samples of regular quartets}
\centering
   \label{tab:distribs}
   \tabcolsep=2pt
\begin{tabular}{r l l  c  c  c  c  c  c  c c c }
\hline
\hline
& \multicolumn{2}{r}{{\textbf{Sample}}} & \CG & \multicolumn{2}{c}{\CB} & &\multicolumn{2}{c}{\CC} & & \multicolumn{2}{c}{\PCF} \\
\cline{5-6}
\cline{8-9}
\cline{11-12}
& \multicolumn{2}{l}{\textbf{Quantity}} & median 
& median &  $p$
& & median &  $p$
& & median & $p$\\
\hline

(1) & $\left\langle R_{ij}\right\rangle$ (kpc) & & $153$ & $478$ & $\boldred{\pmb{<}10^{-6}}$ & & $313$ & $\boldred{\pmb{<}10^{-6}}$ & &$463$& $\boldred{\pmb{<}10^{-6}}$ \\
(2) & $\sigma_v$ (km s$^{-1}$) & &$188$ & $139$ & $\boldblue{8.5 \pmb{\times} 10^{-5}}$ & & $153$ & $\boldblue{0.0072}$ & & 97 & {$\boldblue{\pmb{<}10^{ -6 }}$} \\
(3) & $\log(L_{r,\mathrm{group}}/\mathrm{L_\odot})$ & & $10.99$ & $11.07$  & 
$\boldred{1.2\times 10^{-4}}$ & & $11.01$ & $0.13$ & & $10.94$ & $\boldblue{0.033}$ \\
(4) & $\log(\mathcal{M}_{200\mathrm{c}}  /\mathcal{M}_\odot)$ & & $13.21$ & $13.16$ & $0.22$ & & $13.16$ & $0.23$ & & $12.95$ & $\boldblue{1.58 \times 10^{-4}}$\\
(5) & $\log(\mathcal{M}_\mathrm{VT}/\mathcal{M}_\odot$) & & $12.90$ & $13.05$ &
   $\boldred{0.039}$ & & $12.97$ & $0.18$ & & $12.70$ & $\boldblue{0.041}$ \\
(6) & $t_\mathrm{cr}$ (Gyr) & & $0.78$ & $3.16$ & $\boldred{\pmb{<}10^{ -6 }}$ & & $1.86$ & $\boldred{ \pmb{<}10^{-6}}$ & & $3.90$ & $\boldred{\pmb{<}10^{-6}}$ \\
(7) & $\mathcal{M}_\mathrm{VT}/L_r$\,($\mathcal{M}_\odot/L_\odot$) & & $76$& $100$&  $\boldred{0.041}$ & & $93$ & $0.12$ & & $55$ & $0.18$\\
(8) & $\Delta_\mathrm{BGG-cen}/ \langle R_{ij} \rangle$ & & $0.56$ & $0.47$ &
   $\boldblue{0.0034}$ & & $0.42$ & $\boldblue{ 2\times 10^{-5}}$ & & $0.45$ & 
  $\boldblue{0.011}$ \\
(9) & $\Delta v_\mathrm{BGG}/\sigma_v$ & & $0.54$ & $0.57$ & $0.40$ & & $0.54$ & $0.45$ & & $0.65$ & $0.26$ \\
(10) & $L_{\mathrm{BGG}}/L_{\mathrm{group}}$ & & $0.62$ & $0.51$
& $\boldblue{\pmb{<}10^{ -6 }}$
& & $0.61$ & $0.35$ & & $0.59$ & $0.17$ \\
(11) & $\Delta M_{r12}$ & & $1.17$ & $0.85$ & $\boldblue{1.2\pmb{\times} 10^{-4}}$ & & $1.17$ & $0.50$ & & $1.04$ & $0.17$ \\

\hline
\end{tabular}
  \tablefoot{
  ${\cal M}_{200\mathrm{c}}$ is the virial mass of the group for \PCF{}s and of the parent group for \CG{}s, \CB{}s, and \CC{}s. The significant $p$-values (estimated from one million random shuffles) are highlighted in bold (respectively blue and red for significantly lower and higher values).}
\end{table*}

 \begin{figure}
\centering
 \includegraphics[width=0.94\hsize]{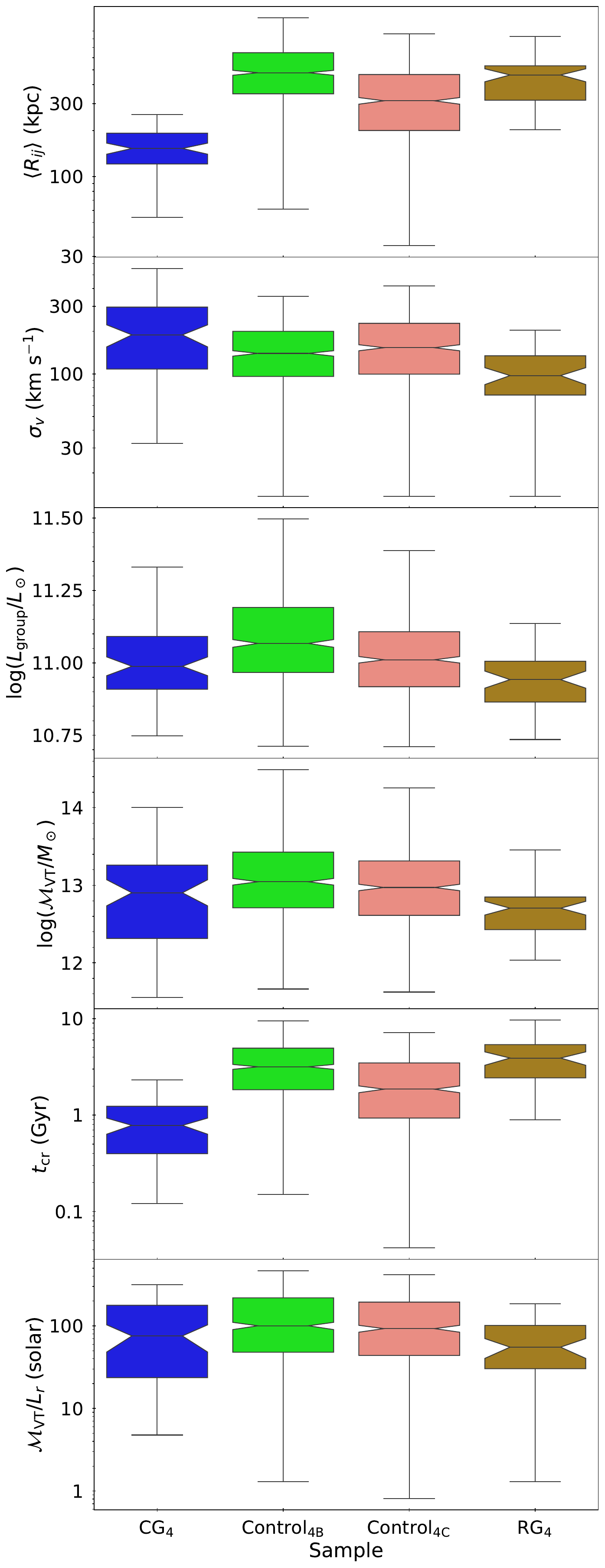}
 \caption{Distributions of group properties for the different samples. From top to bottom, median of the projected inter-galaxy separations, radial velocity dispersion, group $r$-band luminosity, virial theorem mass, crossing time, and virial theorem mass to light ratio. 
The probabilities that the median quantities for the \CB{}, \CC{}, and \PCF\ samples are consistent with those of the \CG{} sample are listed in Table~\ref{tab:distribs}.
\label{fig:Distributions_Groups}
}

\end{figure}

The distributions of these group properties are shown in Table~\ref{tab:distribs}
and displayed for the six most important global parameters in Fig.~\ref{fig:Distributions_Groups}.
Table~\ref{tab:distribs} lists the median quantities and the probabilities that the median quantities of the \CB, \CC, and \PCF\ samples are consistent with those of the \CG\ sample, using random shuffling (where the difference in medians is compared to those obtained in a large number of random sets of same size randomly drawn from the union of the two observed sets).\footnote{Random shuffling has two important advantages over classical non-parametric tests: 1) Shuffling immediately provides probabilities, while the probabilities for classical non-parametric tests are only known for simple (e.g., normal) distributions. 2) Shuffling allows one to compare a particular measure of a distribution, here median, while most classical non-parametric tests measure the consistency between two (full) distributions; thus they are sensitive to measures (e.g., spread) that are not interesting when only testing for a particular measure (e.g., median).}

Given that \CG{}s  are selected to be high mean surface brightness, they should be small and/or luminous. Indeed,
 \CG{}s are much smaller than the groups in the control samples (first row of Table~\ref{tab:distribs} and first panel of Fig.~\ref{fig:Distributions_Groups})
and \CG{}s are also more luminous than \PCF{}s (third row and panel). However,
the \CB{}s are more luminous than the \CG{}s (third row and panel) because their galaxies are selected to be the four most luminous.
There is no difference between the luminosities of \CG{}s and those of the cores of PC groups, i.e. \CC{}s. 

The \CG{} sample has the largest median radial velocity dispersion of the four samples (second row and panel). Given that the \CG{} velocity dispersion notches in Fig.~\ref{fig:Distributions_Groups} do not overlap with the corresponding notches of the control samples, the median of \CG{}s is significantly different than the medians of the control samples ($p<0.007$ according to Table~\ref{tab:distribs}). 
This larger velocity dispersion of \CG{}s  appears to be caused by a too permissive criterion to reject discordant redshifts. Indeed, the criterion of $|v-\mathrm{median}(v)| < 1000\,\kms$ initially introduced by \cite{Hickson+92} is equivalent to a $>5\,\sigma$ rejection criterion, which is much too liberal for redshift space selections. Most authors adopt $3\,\sigma$ rejection criterion, while \cite*{Mamon+10} found that $2.7\,\sigma$ rejection is optimal to recover the velocity dispersions of clusters in cosmological simulations.

Table~\ref{tab:distribs} also provides two measures of group mass: the proxy for the cosmological ``virial" mass, $\mathcal{M}_\mathrm{200\mathrm{c}}$ (fourth row) and the virial theorem masses (Eq.~[\ref{M_VT}], fifth row).
The values of $\mathcal{M}_\mathrm{200c}$ correspond to the total group mass, which are estimated by abundance matching between a known halo mass function and the observed group luminosity function \citep{Lim+17}. On the other hand, the virial theorem masses are measured within the sphere containing the galaxies, using $\mathcal{M}_\mathrm{VT} \propto R\,\sigma_v^2$ (Eq.~[\ref{M_VT}]), which neglects a surface term that leads to an underestimation of the mass in clusters where only an inner subset of galaxies is considered \citep{The&White86}. Indeed, 
in the deeply embedded  \CG{}s (Sect.~\ref{sec:environment} and Fig.~\ref{fig:Virial_offset}), the virial theorem masses are much lower (by 0.31 dex, i.e. a factor of two) than the ``virial" masses, while for the \CB{}  the difference in the logarithms of these two  mass estimates is only 0.11 dex.
Surprisingly, the \CC{} sample of the four closest galaxies, which should also often be Embedded, shows only a 0.19 dex difference in the log mass estimates, perhaps because they are not so embedded since their median sizes are only one-third smaller than for the \CB{}s. It is also interesting to note that the \PCF{} sample shows a high difference of 0.25 dex between these two log mass estimates. While \PCF{}s are extended, their low velocity dispersion leads to low virial theorem mass estimates, while their cosmological virial mass is less affected by their only moderately lower median luminosity.

When comparing the masses of the different group samples, one sees that
both the cosmological virial masses and the virial-theorem masses  of \CG{}s, \CB{}s, and \CC{}s are all higher than those of \PCF{}s, because the former are usually embedded in larger groups (see Sect.~\ref{sec:environment} for the \CG{}s), while the latter, as mentioned above, have unusually low velocity dispersions.
It is interesting to compare the virial theorem mass to luminosity ratios instead of the masses themselves. 
Only the \CB{} subsample displays a significantly different (higher)  median $\mathcal{M}_\mathrm{VT}/L_r$ compared to that for the \CG{}s, because the \CB{}s dominate the \CG{}s more  in $\mathcal{M}_\mathrm{VT}$ (+0.15 dex) than in luminosity (+0.08 dex).

The crossing times of gravitating systems scale as one over square root of the mean density. It is therefore not surprising that \CG{}s, selected to be high surface brightness systems, and which turn out to be mostly small rather than luminous, have 2.5 to nearly 5 times shorter crossing times (Eq.~[\ref{tcr}]) than do the control groups (row 6). 

\subsection{Brightest group galaxies}
\label{sec:BGGs}

We now analyze three properties of the BGGs relative to their group:
1) Their offset in position, 2) their offset in velocity, and 3) their fraction of the group luminosity and offset in absolute magnitude relative to the second most luminous galaxy.

\subsubsection{Spatial offset} \label {sec:offset}
We define the relative offset in terms of the median inter-galaxy projected separation, $\left\langle R_{ij}\right\rangle$. 
Therefore, the relative group spatial offset is  $\Delta_\mathrm{BGG-cen}/\left\langle R_{ij}\right\rangle$, where
$\Delta_\mathrm{BGG-cen}$ is the projected physical separation 
between the BGG and the group centroid projected on the plane of sky.
\begin{figure}
\centering
 \includegraphics[width=0.94\hsize]{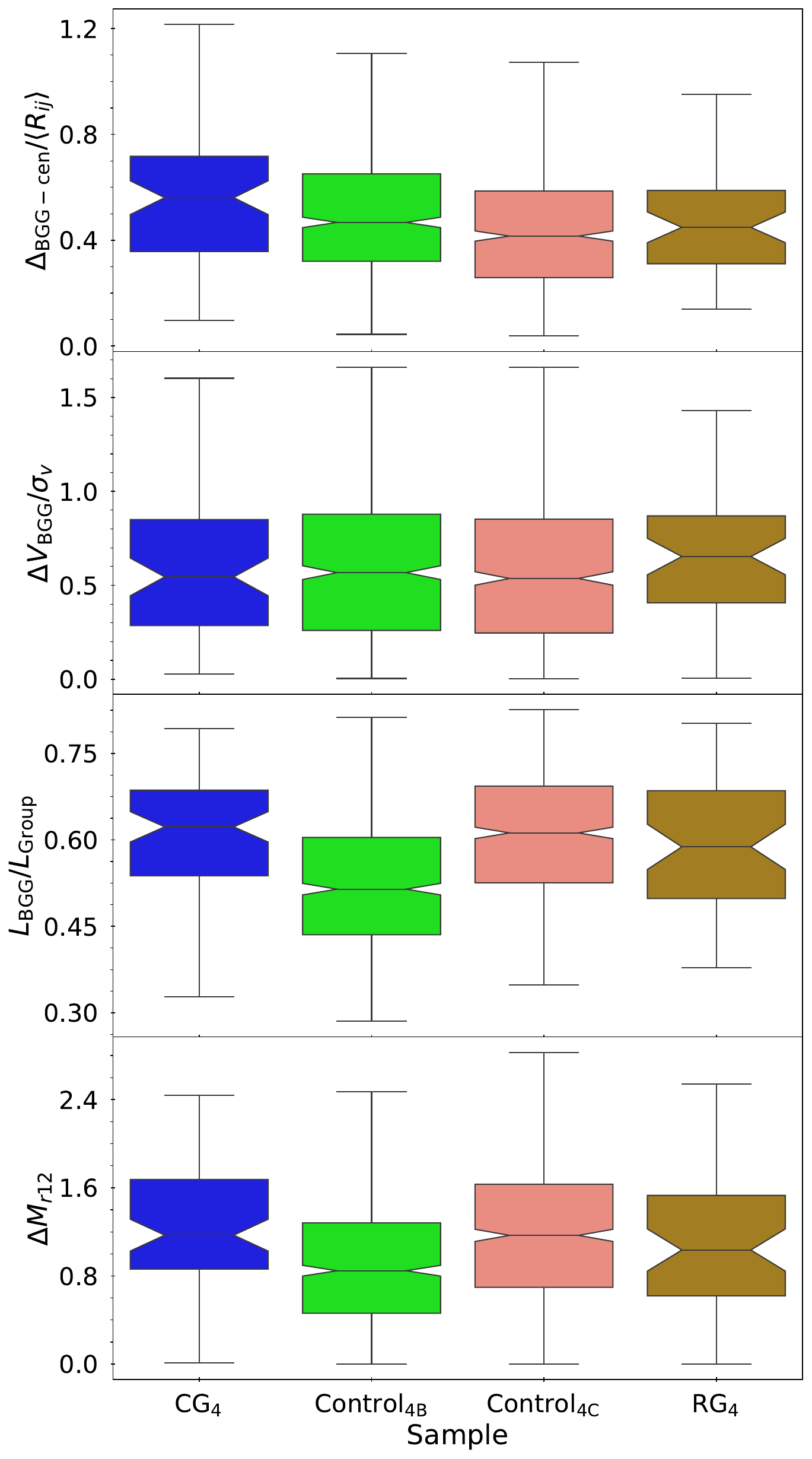}
 \caption{Same as Fig.~\ref{fig:Distributions_Groups} but for  properties related to the BGG.
 From top to bottom, 
 relative BGGs offset from the group geometric center normalized to the median of the inter-galaxy separations,  radial velocity of the BGG relative to the group in units of group velocity dispersion, BGG luminosity fraction, and magnitude gap between the BGG and the second brightest galaxy of the group. 
}
 \label{fig:Distributions_BGG}
\end{figure}
The distributions of these relative offsets are shown in the top panel of Fig.~\ref{fig:Distributions_BGG} and summarized in Table~\ref{tab:distribs}. 

The BGGs in \CG{}s are significantly less centered than the BGGs in the control groups.
As seen in the eighth row of Table~\ref{tab:distribs}, the median relative offsets are $0.56\pm0.03$  for the \CG{} sample, compared to $0.47\pm0.01$  for \CB{}, $0.42\pm0.01$  for \CC{}, and $0.45\pm0.03$ for \PCF{}, where the listed uncertainties are $(p_{84}-p_{16)}/(2\sqrt{2N/\pi}$), with $p_i$ the i$^\mathrm{th}$ percentile and $N$ the sample size.
This could be explained if \CG{}s are plagued by chance projections. 
However, one would then expect that \CG{}s show greater BGG relative velocity offsets compared to in the control samples, but this is not seen (see Sect.~\ref{sec:BGGdv} below).

A simpler explanation is that Lim groups are built around BGGs, so relative to the group centroid, their BGG offsets are low, by construction, while they have no such bias for the BGG velocities.
To be sure, we extracted a subsample of the \cite{Tempel+17} group catalog, which is based on a Friends-of-Friends algorithm, very different from the halo algorithm used to build the \cite{Lim+17} groups.
Our subsample has the same restrictions as used for the \CG{} and Lim groups: same bounds in redshift, absolute $r$-band magnitudes for the brightest and faintest members, as well as restricted to
exactly four members. Those Tempel quartets have a median BGG offset of $0.50\pm0.03$, and the significance of the difference with those of the control groups (estimated through 100\,000 shuffles) are $14\%$ with \CB{} (median of $0.47$), $7\%$ with \CC{} (median of $0.42$) and $1.3\times 10^{-3}$ with \PCF (median of $0.45$). Still, the median BGG offset of 0.56 of the \CG{}s is not significantly higher than that (0.50) of the similarly filtered Tempel quartets ($p=0.48$).

\subsubsection{Velocity offset}
\label{sec:BGGdv}

We computed the relative velocity offsets of the BGGs relative to their host groups as
$\Delta v_\mathrm{BGG}/\sigma_v$, 
which, applying Eq.~(\ref{dv}) to the BGG, is equal to $(z_\mathrm{BGG}-z_\mathrm{group})/ \sigma_z$, where $z_\mathrm{group}$ is the mean redshift of the members and $\sigma_z = (1+z_\mathrm{group})\,\sigma_v/c$ is the redshift dispersion, again
according to Eq.~(\ref{dv}).
In contrast to their projected positions, the BGGs in \CG{}s do not show significantly different relative velocity offsets than those in regular group samples (see  Table~\ref{tab:distribs} and Fig.~\ref{fig:Distributions_BGG}). 

\subsubsection{Dominance of the brightest group galaxy}
\label{sec:BGGdom}
We analyze the dominance of the BGG in two ways: 1) with the fraction of the total group luminosity contained in the BGG; 2) with the magnitude gap, $\Delta M_{r,12} = M_{r,2}-M_{r,1}$, i.e. the difference in the $r$-band rest-frame absolute magnitudes between the BGG and the second brightest galaxy of the group. 

The third panel of Fig.~\ref{fig:Distributions_BGG} shows the fraction of the group luminosity contained in the BGG. 
The BGGs in \CG{}s are as dominant as those in the \CC{} and \PCF{} group samples (the notches overlap). 
\CB{} is the sample showing the least dominant BGGs and Table~\ref{tab:distribs} indicates the  BGG luminosity fraction in \CB{}s is significantly lower than in \CG{}s. This is expected by construction, since we picked the four brightest galaxies in a group, probably competing in luminosity with the BGG.

Similarly, the magnitude gaps of \CG{}s are as wide as those in the \CC{} and \PCF{} group samples, while the \CB\ sample shows significantly lower magnitude gaps.
The median magnitude gaps are 
$1.17\pm0.09$  for \CG{}, $0.85\pm0.03$  for \CB{}, $1.17\pm0.03$  for \CC{}, and $1.04\pm0.11$  for \PCF{} (uncertainties on the medians: $\epsilon(X) = \sqrt{\pi/2}\, \sigma(\Delta M_{r,12})/\sqrt{N}$, where $\sigma$ means standard deviation).

\cite{Tremaine1977}
proposed two statistics,
\begin{equation}
    T_1 =  \frac{\sigma(M_1)}{\left\langle M_2 - M_1\right\rangle}, \quad T_2 =  \frac{1}{\sqrt{0.677}}\frac{\sigma(M_2 - M_1)}{\left\langle M_2 - M_1\right\rangle} \ ,
    \label{T1T2}
\end{equation}
to test the importance of magnitude gaps. They
 proved that $T_1$ and $T_2$ should both be greater than unity  for cumulative luminosity functions that are single or double power-laws. 
Galaxy mergers tend to grow the most massive galaxy at the expense of the second-rank galaxy (\citealt{Mamon87} for N-body simulations of virialized groups and \citealt{Farhang+17} for groups in a cosmological context using SAMs), causing $T_1$ and $T_2$ to rapidly fall below unity \citep{Mamon87}.
 
We computed $T_1$ and $T_2$ for our four samples. Table~\ref{tab:Tremaine} shows very low values of $T_2$ and especially $T_1$ for all four group samples. While values below unity were previously found for clusters \citep{Tremaine1977} and the 2MASS compact groups \citep{DiazGimenez+12}, they were never found before for non-compact poor groups.
This suggests that all four group samples have overluminous BGGs, most probably caused by mergers.

Tremaine-Richstone statistics samples built from other catalogs of compact and regular groups are displayed in the lower half of Table~\ref{tab:Tremaine}. These statistics are computed for subsamples filtered with the following criteria: 1) exactly four members; 2) the same maximum BGG redshift; 3) the same minimum BGG luminosity; 4) range in absolute magnitudes $M_{r,4}-M_{r,1}\leq 3$.

\begin{table}
  \caption{Tremaine-Richstone statistics for different group samples with identical selection criteria.}
    \centering
    \begin{tabular}{lrcc}
        \hline
        \hline
Sample & \multicolumn{1}{c}{Size} & $T_1$ & $T_2$ \\        \hline
        \textbf{\CG} & $\mathbf{78}$ & $\mathbf{0.34 \pm 0.03}$ & $\mathbf{0.60 \pm 0.05}$  \\
        \CB & $699$ & $0.37 \pm 0.01$ & $0.78 \pm 0.02$  \\
        \CC & $704$ & $0.29 \pm 0.01$ & $0.65 \pm 0.02$  \\
        \PCF & $56$ & $0.19 \pm 0.02$ & $0.76 \pm 0.06$  \\
        \hline
        D12 & $59$ & $0.51\pm0.06$ & $0.61\pm0.06$ \\
        S16 & $31$ & $0.35\pm0.05$ & $0.84\pm0.10$ \\
        Z20 & $14$ & $0.16\pm0.05$ & $0.80\pm0.18$ \\
        D18 & $117$ & $0.35\pm0.03$ & $0.66\pm0.05$ \\
        Z22 & $88$ & $0.37\pm0.04$ & $0.72\pm0.06$ \\
        \hline
        L17 & $69$ & $0.19\pm0.02$ & $0.75\pm0.06$ \\
        T21 & $43$ & $0.12\pm0.02$ & $0.65\pm0.08$ \\
        \hline
        \end{tabular}
    \label{tab:Tremaine}
\tablefoot{ 
    All samples are in fact subsamples of four galaxies having $M_\mathrm{BGG}\leq -21.81$  and fewer than three magnitudes between the brightest and the faintest, with all galaxy redshifts between 0.005 and 0.0452.
    The errors were estimated from 10\,000 bootstraps. The first five samples (D12: \citealt{DiazGimenez+12}; S16: \citealt{Sohn+16}; Z20: \citealt{Zheng&Shen20}; D18: \citealt{DiazGimenez+18}; Z22: \citealt{Zandivarez+22} are designed to be compact, while the latter two (L17: \citealt{Lim+17}; T21: \citealt{Tinker21}) are not. The \CG{}s are selected from D18, but in the somewhat  smaller \LT\ region.
    }
\end{table}

Table~\ref{tab:Tremaine} indicates that all group samples besides D12 (the 2MCG sample from \citealt{DiazGimenez+12}) have $T_1$ much lower than unity ($T_1\leq 0.4$). The $T_1$ values for the original HCG sample was 1.16 \cite{Mamon86}, but this was due to a bias of the HCG sample against systems very dominated by a single galaxy (despite the magnitude concordance criterion, see \citealt{DiazGimenez&Mamon10}). For their full 2MCG sample, D12 found  $T_1=0.51\pm0.06$, precisely what is given in Table~\ref{tab:Tremaine}  for the ``D12" subsample of 2MCG. 
The lower $T_1$ values here appear to be the consequence of using volume-limited samples (restriction in redshift and absolute magnitude of the BGG that we imposed on the samples).

\begin{table}
\caption{Tremaine-Richstone statistics of samples of \LT\ groups with different selections.
\label{tab:TremaineVarious}} 
\centering
\tabcolsep=4pt
\begin{tabular}{lcrcccc}
\hline\hline
$N_\mathrm{max}$ & Selection & \multicolumn{1}{c}{Size} & $T_1$ & $T_2$ & $\langle M_2 - M_1\rangle$ & $\sigma(M_1)$  \\
\hline
\ 4+ & $r$ 
    & $2858$ & $0.68$ & $0.83$ & $1.05$ & $0.71$\\
\ 4+ & $M_r, z$ 
& $1038$ & $0.30$ & $0.82$ & $1.14$ & $0.34$ \\
\ 4 & $r$ & $620$ & $0.54$ & $0.75$ & $1.36$ & $0.73$\\
\ 4 & $M_r, z$ & $200$ & $0.16$ & $0.69$ & $1.53$ & $0.25$\\
\hline
\end{tabular}
\tablefoot{Here, 4+ (resp. 4) stands for at least (resp. exactly) four members in the group. Selection criteria are the following: $r$ means $r<17.77$ (SDSS Main Galaxy Sample) and $r_\mathrm{BGG} < 14.77$; $M_r, z$ means $0.005 < z_\mathrm{BGG} < 0.0452$ and $M_{r,\mathrm{BGG}} < -21.81$.}
\end{table}

We used
\LT\ groups to test if our limits on group multiplicity on one hand and on redshift and luminosity range on the other could be the cause of our lower $T_1$ values. 
As seen in Table~\ref{tab:TremaineVarious}, the imposition of a volume- and luminosity-limited sample for the BGGs drastically reduces the values of $T_1$ because of the drastic reduction of the spread of BGG magnitudes with a small rise in the mean magnitude gap.
Indeed, as seen in Fig.~\ref{fig:CG_Selection}, going from the apparent magnitude cut for both the centrals and the satellites (oblique lines and oblique edge of shaded region)  to  the luminosity cut for both (horizontal lines) discards the groups with BGG luminosities below the cut, without affecting much the magnitude gap.

\subsection{\CG{} properties according to their location within their host \LT\ groups}

\begin{table*}  
\caption{Comparison of median properties of \CG{} sub-samples according to their location in \LT\ groups \label{tab:CGLT}}
\centering
   \tabcolsep=3pt
\begin{tabular}{r l c c  c c c  c c c c  c  c  c c}
\hline
\hline
& \multicolumn{1}{r}{{\bf Sample}} & 
\multicolumn{2}{c}{Split} & & 
\multicolumn{3}{c}{Isolated} & & 
\multicolumn{2}{c}{Embedded} & & 
\multicolumn{2}{c}{Predominant} \\
\cline{3-4}
\cline{6-8}
\cline{10-11}
\cline{13-14}
& {\bf Quantity} & median  &  $p_{\CG}$ & & median &  $p_{\CG}$&  $p_{\PCF}$ & & median &  $p_{\CG}$ & & median & $p_{\CG}$\\
\hline
(1) & Number & \multicolumn{2}{c}{$16$} & & \multicolumn{3}{c}{$6$} & & \multicolumn{2}{c}{$19$} & & \multicolumn{2}{c}{$37$} \\
(2) & Common BGG & \multicolumn{2}{c}{$4$ ($0.25$)} & & \multicolumn{3}{c}{6 (1.00)} & &  \multicolumn{2}{c}{$13$ ($0.68$)} & & \multicolumn{2}{c}{37 (1.00)} \\
(3) & $\left\langle R_{ij}\right\rangle$ (kpc) &  $145$ & $0.28$ & & $171$ & $0.26$ & $\boldblue{4.0\pmb{\times}10^{-6}}$ & & $153$ & $0.49$& & $158$\ & $0.37$ \\
(4) &$\sigma_v$ (km s$^{-1}$) &  $366$ & $\boldred{0.0011}$ & & $68$ & $\boldblue{7.5\times 10^{-4}}$ & $0.084$ & & $191$ & $0.37$ & & $158$ & $0.14$ \\
(5) &$\log(L_{r,\mathrm{\CG}}/\mathrm{L_\odot})$ &  $10.98$ & $0.41$ & & $10.92$ & $0.12$ & $0.39$ & & $11.09$ & $\boldred{0.025}$& & $10.98$ & $0.36$ \\
(6) &$\log(\mathcal{M}_\mathrm{VT}/M_\odot)$ & $13.33$ & $\boldred{0.0032}$ & & $12.03$ & $\boldblue{0.0079}$ & $\boldblue{9.9 \times 10^{-4}}$& & $12.95$ & $0.40$ & & $12.55$ & $0.10$\\
(7) &$t_\mathrm{cr}$ (Gyr) &  $0.38$ & $\boldblue{0.0026}$& & $1.8$ & $\boldred{0.0071}$ &$\boldred{0.011}$& & $0.67$ & $0.29$ & & $0.98$ & $0.10$ \\
(8) &$\log(\mathcal{M}_\mathrm{200c}/M_\odot$) & -- & -- & & $12.86$ &$\boldblue{0.022}$& $0.15$ & & $13.83$ & $\boldred{0.0015}$ & & $13.13$ & $0.20$ \\
(9) &$\mathcal{M}_\mathrm{VT}/L_r$\,($\mathcal{M}_\odot/L_\odot$)&  $241$ & $\boldred{0.0020}$& & $13.3$ & $\boldblue{0.024}$ & $\boldblue{0.0011}$& & $53$ & $0.42$ & & $37$ & $0.12$ \\
(10) &$\Delta_\mathrm{BGG-cen}/ \langle R_{ij} \rangle$ &  $0.55$ & $0.48$ & &$0.58$ & $0.36$ & $0.08$ & & $0.52$ & $0.40$ & & $0.55$ & $0.49$ \\
(11) &$\Delta v_\mathrm{BGG}/\sigma_v$ &  $0.54$ & $0.47$ & & $0.61$ &$0.45$ & 0.43 & & $0.69$ & $0.31$ & & $0.51$ & $0.41$ \\
(12) &$L_{\mathrm{BGG}}/L_{\mathrm{\CG}}$ &  $0.60$ &$0.34$ & & $0.52$ & $\boldblue{0.047}$ & $0.12$ & & $0.65$ & $0.14$  & & $0.63$ & $0.42$ \\
(13) &$\Delta M_{r12}$ & $1.15$ & $0.47$ & & $0.65$ & $\boldblue{0.035}$ & $0.09$ & & $1.32$ & $0.26$ & & $1.14$ & $0.47$ \\
\hline
\end{tabular}
\tablefoot{The four CG classes are those of \cite{Zheng&Shen21}. The values in the second row are the number of groups with the corresponding fractions  in parentheses. Column titles
$p_\textbf{\CG}$ and $p_\textbf{\PCF}$ are the $p$-values of the difference of those medians with \CG\ and \PCF, respectively. They are both estimated through $10^6$ random shuffles. Significant differences are displayed in bold, in red or blue for significantly higher or lower values compared to the reference value, respectively. Row (2) displays the number and fraction of \CG{}s of the given class. In row (8), $\mathcal{M}_\mathrm{200c}$ is relative to the parent group of considered subgroups. We cannot compute $\mathcal{M}_\mathrm{200c}$ for Split \CG{}s.
}
\end{table*}

In Sect.~\ref{sec:locvsZheng}, we divided the CG4s into four \citeauthor{Zheng&Shen21} classes: Split, Isolated, Predominant, and Embedded. We now analyze the possible dependencies of the properties of \CG{}s on the environment they inhabit. 
The properties of the four sub-samples of \CG{}s are displayed in Table~\ref{tab:CGLT}.

The Split \CG{}s have significantly higher median velocity dispersion ($366\,\kms$) than the full set of \CG{}s ($188\,\kms$). In fact, the non-Split \CG{}s have a median velocity dispersion of only $160\,\kms$. However, the median velocity dispersion of the \PCF{}s is much lower at $\sigma_v = 97\, \kms$ ($p=10^{-5}$).  
This confirms our long-held suspicion that the velocity concordance criterion that all member galaxies must lie within $\pm1000\,\kms$ from the median is too permissive, since it is over five times the median velocity dispersion of the full set of \CG{}s (Table~\ref{tab:distribs}) and roughly seven times the median velocity dispersion of non-Split \CG{}s. Filtering our \CG{} sample to those whose velocities all lie within $500\,\kms$ from the median (instead of $1000\,\kms$), we find only 7 Split compact groups among 64, i.e. only 11\% (instead of 25\%).\footnote{Actually, among 78 \CG{}s, three have  $\sigma_v \ge 500 \kms$ (the maximum \CG{} velocity dispersion is $686 \kms$),  all of which are Split groups. In fact, the five highest velocity dispersion \CG{}s are all Split.}
Therefore, the Split \CG{}s may be contaminated by galaxies that are chance-aligned with at least another group). This is corroborated by the low fraction of common BGGs between Split groups and their two or more host groups (row 2 of Table~\ref{tab:CGLT}). This higher velocity dispersion of Split \CG{}s compared to other \CG{}s, at equal size, translates to higher virial theorem mass and  mass-to-light ratio, as well as lower crossing time (rows 6 to 8 of Table~\ref{tab:CGLT}) for the Split \CG{}s. 

The Isolated \CG{}s display a large number of statistically significant (bold in Table~\ref{tab:distribs}) differences with the full set of \CG{}s: they have  lower velocity dispersions, virial theorem masses, and magnitude gaps, as well as higher crossing times. 
The lower velocity dispersions of the Isolated \CG{}s compared to the Predominant and Embedded ones are probably the consequence of the higher multiplicities of the parent groups of these latter two classes of \CG{}s. These host groups may have galaxies outside of the \CG{}s on the plane of sky that are physically near their BGG (in real space), implying more mass at this physical distance to the BGG and hence a higher velocity dispersion. In turn, the lower velocity dispersions of Isolated CG{}s produce lower virial theorem masses and higher crossing times by the definitions of these quantities.
Finally, the Isolated \CG{}s have significantly  lower magnitude gaps (and marginally significantly lower BGG luminosity fractions) than the ensemble of \CG{}s. 
This is expected, since the Embedded and Predominant \CG{}s  usually share the BGG of their host group (second row of Table~\ref{tab:CGLT}), which is usually richer, more massive  and with more luminous BGGs (Table~\ref{tab:distribs}).

We also note that the Isolated \CG{}s have indistinguishable median properties compared to the \PCF{}s. Only CG{} properties that depend on group size, namely size, virial theorem mass, crossing time, and virial theorem mass-to-light ratio have significantly lower values than in the \PCF{}s.

The Embedded \CG{}s have higher luminosity than the typical \CG, as expected by their definition of contributing over half the luminosity of the host group. 
Finally, the Predominant \CG{}s 
show no significant differences with the full sample, perhaps because they account for half the \CG{} sample.

Row 8 does not display $\mathcal{M}_\mathrm{200c}$ for the parent group of Split groups, because there is more than one parent. Predominant parent groups have the typical \CG{} $\mathcal{M}_\mathrm{200c}$, while Isolated and Embedded ones are respectively more and less massive.
Indeed, with exactly four members, 
\CG{}s lie in a narrow mass span compared to their parent groups. Therefore,  Isolated parent groups are the \CG{} itself and therefore less massive, while Embedded parent groups have to be large enough so that the considered \CG{} is not Predominant.

\subsection{Correlations of group properties}

\subsubsection{Luminosity segregation}
\label{sec:lseg}
\begin{figure}
\centering
 \includegraphics[width=0.9\columnwidth]{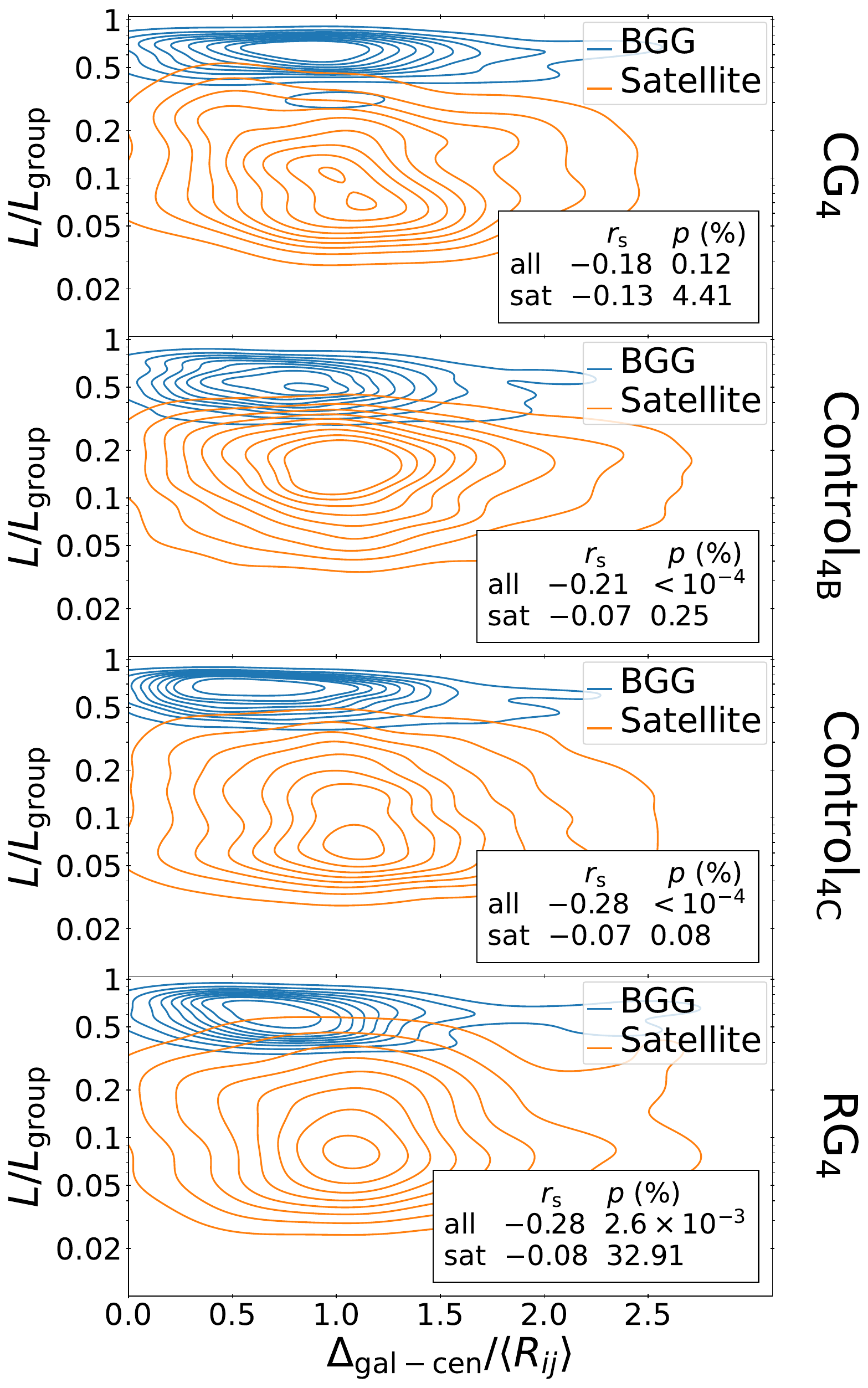}
 \caption{Luminosity segregation in the four group samples: galaxy luminosity fraction versus relative position of galaxy relative to the group centroid.
The panels show the kernel density estimator contour level lines for \CG, \CB{}, \CC{}, and \PCF{} from top to bottom, respectively. The BGGs are shown in blue and the satellites in orange. 
The rank correlations and corresponding $p$ values (in percent) are displayed for all galaxies of the group (i.e., including BGGs) and for the satellite galaxies (i.e., non-BGGs) only.
\label{fig:Luminosity_segregation}
}
\end{figure}

In relaxed galaxy systems, (partial) energy equipartition should lead to mass segregation, where the most massive galaxies  lie in the group center, while the low-mass ones will be able to probe the outskirts of the group. If luminosity traces mass, one would thus expect luminosity segregation with an anticorrelation between galaxy luminosity and distance to the group center.
Luminosity segregation was predicted for CGs by \cite{Mamon87} and first detected in CGs by \cite{DiazGimenez+12}, who saw no signs of it in other regular group samples\footnote{\cite{DiazGimenez+12} did not analyze the \cite{Yang+05} group sample, which resembles the \cite{Lim+17} sample.}, nor in the original CG sample by \cite{Hickson+92}, which is incomplete in systems with a dominant BGG \citep{DiazGimenez&Mamon10}.

Figure~\ref{fig:Luminosity_segregation} displays the trend of galaxy luminosity fraction versus relative position in the group.
There exists a luminosity segregation for \CG{} that is also present in the control samples. The Spearman rank test of luminosity fraction of galaxies versus distance to centroid in median inter-distance units produces  a significant anticorrelation  ($r_\mathrm{S}=-0.18$, with a probability $p=0.01$ of occurring by chance). The control samples show even stronger luminosity - radial distance anticorrelations. 

Given that BGGs are understood to grow by mergers (\citealt{Mamon87}), we also searched for luminosity segregation restricted to the satellites.
Figure~\ref{fig:Luminosity_segregation} shows that luminosity segregation among satellites is still detected in \CG{}s as well as in the control samples, but with far weaker (except in \CG{}s), yet still significant, correlations  (except for \PCF{}s).

We now consider luminosity segregation for the different \CG{} \citeauthor{Zheng&Shen21} classes.
Isolated satellites show a  significant luminosity segregation: $r_\mathrm{S}=-0.48$ ($p=0.04$), despite containing only six groups. Surprisingly, the luminosity segregation is decreased and loses its statistical significance when adding the BGGs 
($r_\mathrm{S} = -0.20$, $p = 0.36$).
Interestingly, the Split \CG{}s are also the sites of significant luminosity segregation: 
$r_\mathrm{S}=-0.24$ ($p=0.05$) for all member galaxies and $r_\mathrm{S}=-0.27$ ($p=0.007$) when restricted to satellites.
The other two \CG{} subsamples do not show any significant luminosity segregation, whether for all galaxies or for just the satellites, except for a marginally significant segregation ($r_\mathrm{S} = -0.15$, $p = 0.07$) when all galaxies from Embedded groups are considered.

\subsubsection{Other significant correlations within samples}

\begin{table}
\caption{Significant correlations between group properties. \label{tab:corr}}
\centering
  \tabcolsep=3pt
  \begin{tabular}{lllrr}
    \hline
        \hline
   Param-1    & Param-2 & Sample & Corr. & Prob.\\ 
           (1) & (2) & (3) & (4) & (5) \\
    \hline
    \rule{0pt}{3ex} 
            $\left\langle R_{ij}\right\rangle$ (kpc) & $\sigma_v$ (km s$^{-1}$) &  Control$_{4\mathrm{C}}$ & $-0.30$ & $< 10^{-6}$ \\
\rule{0pt}{3ex}  
             $\left\langle R_{ij}\right\rangle$ (kpc) & $\Delta_\mathrm{BGG-cen}/ \langle R_{ij} \rangle$ & CG$_4$ & $-0.26$ & $2.4 \times 10^{-2}$ \\
             &  & Control$_\mathrm{4\mathrm{B}}$ & $-0.07$ & $5.5 \times 10^{-2}$ \\
\rule{0pt}{3ex}         
            $\sigma_v$ (km s$^{-1}$) & $\log(L_\mathrm{group}/L_\odot)$ & Control$_{4\mathrm{C}}$ & $0.34$ & $< 10^{-6}$ \\
             &  & Control$_\mathrm{4\mathrm{B}}$ & $0.33$ & $< 10^{-6}$ \\
\rule{0pt}{3ex}  
             $\sigma_v$ (km s$^{-1}$) & $\Delta v_\mathrm{BGG}/\sigma_v$ & Control$_{4\mathrm{C}}$ & $-0.14$ & $1.3 \times 10^{-4}$ \\
             &  & Control$_\mathrm{4\mathrm{B}}$ & $-0.10$ & $7.0 \times 10^{-3}$ \\
\rule{0pt}{3ex}  
            $\sigma_v$ (km s$^{-1}$) & $L_{\mathrm{BGG}}/L_{\mathrm{Group}}$ &  Control$_\mathrm{4\mathrm{B}}$ & $-0.13$ & $5.4 \times 10^{-4}$ \\
             &  & Control$_{4\mathrm{C}}$ & $0.10$ & $6.9 \times 10^{-3}$ \\
\rule{0pt}{3ex}  
             $\sigma_v$ (km s$^{-1}$) & $\Delta M_{r,12}$& Control$_{4\mathrm{C}}$ & $0.10$ & $9.7 \times 10^{-3}$ \\
             &  & Control$_\mathrm{4\mathrm{B}}$ & $-0.08$ & $3.4 \times 10^{-2}$ \\
\rule{0pt}{3ex}  
            $\log(L_\mathrm{group}/L_\odot)$ & $\Delta V_\mathrm{BGG}/\sigma_v$ & Control$_\mathrm{4\mathrm{B}}$ & $0.07$ & $4.9 \times 10^{-2}$ \\
\rule{0pt}{3ex}  
            $\Delta_\mathrm{BGG-cen}/ \langle R_{ij} \rangle$ & $\Delta M_{r,12}$ & Control$_\mathrm{4\mathrm{B}}$ & $-0.18$ & $1.0 \times 10^{-6}$ \\
             &  & Control$_{4\mathrm{C}}$ & $-0.14$ & $1.2 \times 10^{-4}$ \\
\rule{0pt}{3ex} 
            $\Delta_\mathrm{BGG-cen}/ \langle R_{ij} \rangle$ & $L_{\mathrm{BGG}}/L_{\mathrm{Group}}$ & Control$_\mathrm{4\mathrm{B}}$ & $-0.18$ & $2.2 \times 10^{-6}$ \\
             &  & Control$_{4\mathrm{C}}$ & $-0.16$ & $1.9 \times 10^{-5}$ \\
\rule{0pt}{3ex}  
             $\Delta v_\mathrm{BGG}/\sigma_v$ & $L_{\mathrm{BGG}}/L_{\mathrm{Group}}$ & Control$_{4\mathrm{C}}$ & $-0.09$ & $2.3 \times 10^{-2}$ \\
             &  & Control$_\mathrm{4\mathrm{B}}$ & $-0.08$ & $3.9 \times 10^{-2}$ \\
\rule{0pt}{3ex}  
            $\Delta v_\mathrm{BGG}/\sigma_v$ & $\Delta M_{r,12}$& Control$_{4\mathrm{C}}$ & $-0.10$ & $1.1 \times 10^{-2}$ \\
             &  & Control$_\mathrm{4\mathrm{B}}$ & $-0.09$ & $2.0 \times 10^{-2}$ \\
\hline 
    \end{tabular}
\tablefoot{Columns: (1): first parameter; (2); second parameter; (3): sample; (4): Spearman rank correlation coefficient; (5): probability of trend for uncorrelated samples. 
  Several pairs of parameters for which selection effects should be strong have been omitted. 
  }
\end{table}

Next, we searched for parameter correlations other than luminosity segregation.
We excluded correlations that are caused by selection effects on \CG{}s:
\begin{itemize}
\item Magnitude gap versus BGG luminosity fraction. They both measure the dominance of the BGG, relative to the second-ranked galaxy and to the group itself, respectively.
\item Group luminosity versus group size since compact groups are defined through a minimal surface brightness.
    \item Group luminosity versus BGG luminosity fraction, because the product of the group luminosity and BGG luminosity fraction is the BGG luminosity, for which we set a lower limit (see Sect.~\ref{sec:data}).
    \item Group luminosity versus magnitude gap since this parameter is correlated to the BGG luminosity fraction (see previous point).
\end{itemize}

Table~\ref{tab:corr} displays the significant correlations seen in the different group samples between group size, velocity dispersion, luminosity, BGG positional and velocity offsets, as well as luminosity fraction and magnitude gap, after discarding the correlations noted in the previous paragraph.
\begin{figure*}
\centering
 \includegraphics[width=\hsize]{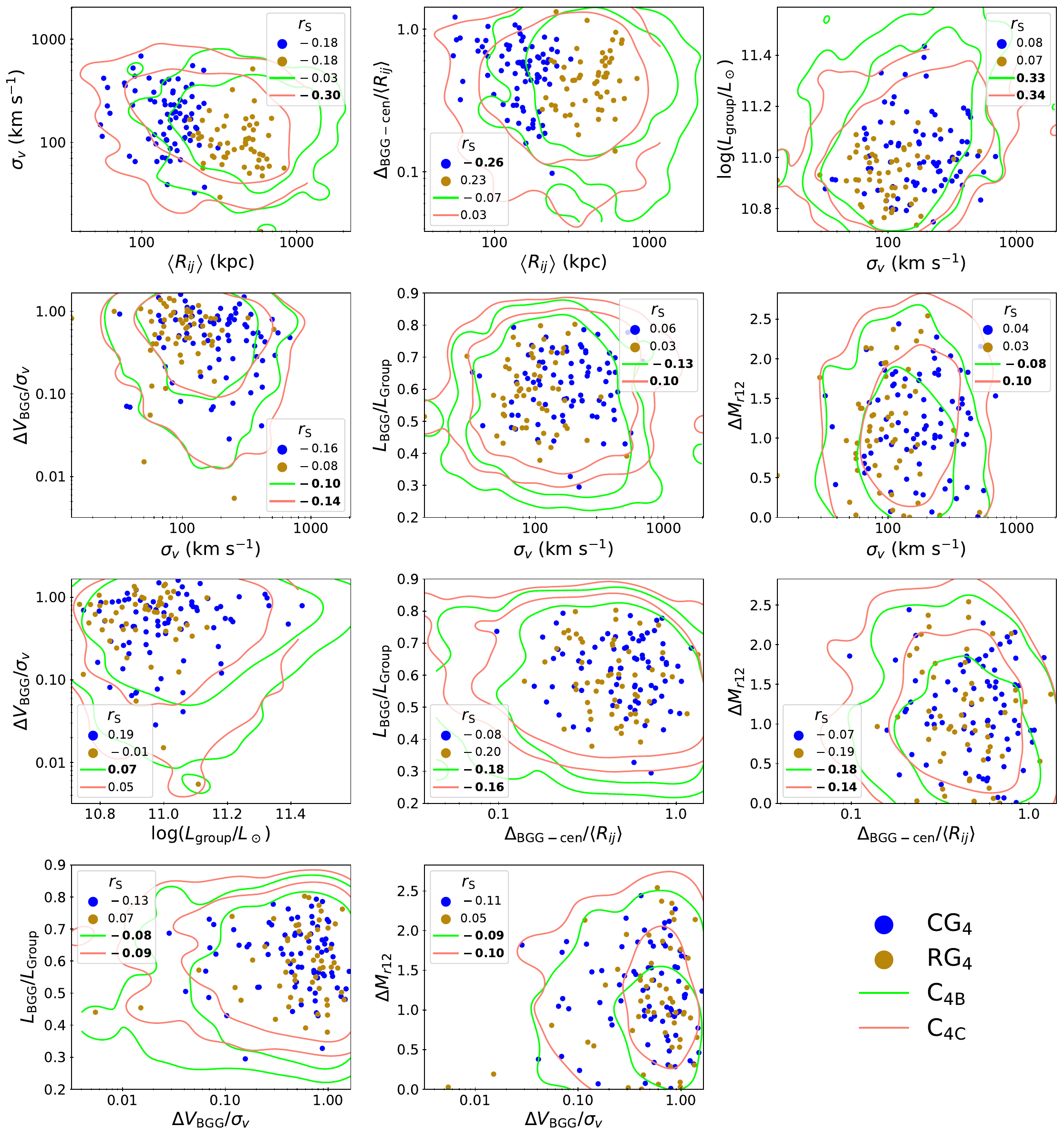}
 \caption{Most significant correlations between quantities within samples as shown in Table~\ref{tab:corr}. Quantities strongly correlated because of selection effects are not displayed - see main text. The legends provide the Spearman rank correlations, which are in bold if significant.
}
 \label{fig:Sig_corr}
\end{figure*}
The corresponding plots for all pairs of quantities showing significant correlations for at least one sample are displayed in Fig.~\ref{fig:Sig_corr}. 

The \CG\ groups show only one significant correlation: group size (from the median projected separation) versus
BGG relative offset ($r_\mathrm{S}=-0.26$, $p=0.024$).
 
The other group samples show some significant correlations not seen in \CG{}s. The most significant of these are the anticorrelations of relative BGG offset with the dominance of the BGG (luminosity fraction of BGG or magnitude gap). This is an improvement over \cite{Skibba+11} who showed that the BGG does not lie at the geometric center of groups, and complementary to \cite{Gozaliasl+19} who showed that the BGG offset decreases with increasing halo mass, decreasing redshift and increasing magnitude gap. 
Also, the velocity dispersions of both \CB{}s and \CC{}s are correlated with their total luminosities. 
These velocity dispersions are even more correlated with the group masses, as expected from the virial theorem mass definition incorporating the group velocity dispersion, and from the Yang-Lim group finder algorithm that uses mass to predict velocity dispersion and deduce membership.
Interestingly, the velocity dispersions of \CC{}s are correlated with their BGG luminosity fraction, while the opposite trend is significant for the \CB{}s.
We found no significant correlations of properties for the \PCF{} sample, because of its much smaller size (as that of the \CG{} sample) compared to those of the \CB{}s and \CC{}s.

\section{Conclusions and final discussion}
\label{sec:concdisc}
\subsection{Conclusions}
\label{sec:discus}
We have here compared a doubly-complete sample of compact groups, restricted to groups of four members to avoid multiplicity effects, to three carefully crafted control samples of regular groups. One consists of the groups of at least four members, from which we select the brightest group galaxy and its three closest satellites. Another consists of the groups of at least four members, from which we select instead the brightest group galaxy and its three brightest satellites. And the last one consists of regular groups of exactly four galaxies.  This comparison led to the following conclusions.
\begin{enumerate}
 \item  A large majority of the \CG{}s are located in the cores of their host groups, and a vast majority of those share their BGG with their host group (Fig.~\ref{fig:Virial_offset}). 
\item Only a small fraction (8\%) of the \CG{}s are identical to a Lim
  group (Sect.~\ref{sec:locvsZheng}). Furthermore, 10\% of isolated regular Lim groups of four members are \CG{}s.
 \item The \CG{}s are smaller, as expected from their selection (Table~\ref{tab:distribs} and Fig.~\ref{fig:Distributions_Groups}).
\item The \CG{}s have higher velocity dispersions than ordinary groups (Table~\ref{tab:distribs} and Fig.~\ref{fig:Distributions_Groups}).
This suggests that the velocity concordance criterion of $<1000\,\kms$ from the median for \CG{}s is too liberal, as it amounts to over $5\,\sigma_v$ for the full set of \CG{}s and nearly $7\,\sigma_v$ for those that are not split between two or more parent groups. Had we taken a $2.7\,\sigma$ rejection criterion \citep{Mamon+10}, the maximum allowed difference with the median velocity would have been $500\,\kms$ when including the Split \CG{}s and $400\,\kms$ when discarding them 
(but see item \#\ref{CGSplit} below).
Furthermore, \cite{Zandivarez+24} found that CGs selected with a velocity concordance of $500\,\kms$ have a 20\% lower median velocity dispersion than when using the usual $1000\,\kms$ velocity concordance limit. 
Reducing the velocity dispersion of \CG{}s by 20\% would make the median velocity dispersion match
that of \CC{}.

 \item The \CG{}s are less luminous than the \CB{}s (which is not surprising) and marginally less luminous than the \CC{}s but more luminous than the \PCF{}s (Table~\ref{tab:distribs} and Fig.~\ref{fig:Distributions_Groups}).
 \item The BGGs in \CG{}s have stronger relative spatial offsets than those in ordinary \LT{} groups (Table~\ref{tab:distribs} and Fig.~\ref{fig:Distributions_BGG}). But those control groups are built according to the \cite{Yang+07} group finder updated by \cite{Lim+17}, which builds the group around the BGG. Comparing instead to a similarly filtered sample of \cite{Tempel+17} groups of four galaxies shows no significant difference in BGG offsets (end of Sect.~\ref{sec:offset}). 
 
 \item  The BGGs in \CG{}s contribute to similar fractions of the group luminosity as do the control samples, except for the \CB{}s (Table~\ref{tab:distribs} and Fig.~\ref{fig:Distributions_BGG}), which are designed to host the four most luminous galaxies and are more likely to have similar absolute magnitudes instead of having a dominant one.
Similarly, \CG{}s have similar magnitude gaps between the first- and second-ranked luminosity galaxies, as do the control groups, except for \CB{}s, which exhibit a smaller magnitude gap for the same reasons as above. 
\item 
The \CG{}s show significant luminosity segregation, as do the three control group samples (Fig.~\ref{fig:Luminosity_segregation}).
And even after removing the BGG, the remaining galaxies (i.e., satellites) show a significant luminosity segregation, and it is stronger than in the three control samples. 
\item Only 8\% of the \CG{}s, selected to be Isolated in redshift space are not associated with larger groups (Sect.~\ref{sec:locvsZheng}), in contrast to 27\% for the \cite{Zheng&Shen20} CGs of at least three galaxies \citep{Zheng&Shen21}, as expected from the decreasing multiplicity function of cosmic systems. Those Isolated \CG s are smaller (as expected from selection effects) and less massive than a typical non-compact group of four galaxies (i.e., an \PCF).

\item \label{CGSplit}
One quarter of the \CG{}s are split between several regular groups (Sect.~\ref{sec:locvsZheng}), which is the same fraction \cite{Zheng&Shen21} found for their CGs.
The Split \CG{}s have much larger velocity dispersions (Table~\ref{tab:CGLT}), suggesting that this subsample may be contaminated by spurious CGs caused by chance alignments of galaxies within unconnected groups.
The Split \CG{}s also show significant luminosity segregation for all galaxies, even when  discarding the BGGs and thus restricting Split \CG{}s to the satellites. But this luminosity segregation may be spurious (Sect.~\ref{sec:lseg}). For example, it may be caused by the superposition of regular groups of different characteristics along the line of sight.
\end{enumerate}

In summary, apart from the properties directly affected by selection effects (high mean surface brightness, implying smaller and more luminous groups), the properties of  compact groups of four galaxies selected to be truly isolated in redshift space (i.e., of the Isolated class) are quite similar to those of regular groups of galaxies. Hence, compact groups are not special.
Our conclusion (based on observations) is both similar and complementary to that of \cite{Zandivarez+14}, who compared the galaxy locations in mock CGs to mock regular groups in a SAM after supplementing the CGs with galaxies less luminous than the three-magnitude range limit. They found very similar galaxy surface density profiles in terms of the projected distance of ``satellites" to the first- or second-ranked galaxy, normalized to the projected distance between these two brightest galaxies.

\subsection{Particularities of \CG{}s and clues to their nature}

Our results provide clues on the nature of compact groups of galaxies. 
Since the great majority of the \CG{} BGGs are common to those of parent group BGGs, it is tempting to consider such compact groups as physically dense cores of parent groups. More precisely, \CG{}s would constitute the cores of regular groups that are fortunate to appear isolated along the line of sight. 

However, having associations inside a parent group does not guarantee that these groups are physically dense. Indeed, it is common to have chance alignments along the line of sight of galaxies -- or better pairs of galaxies -- within groups, as predicted by \cite{Walke&Mamon89}. This has been  confirmed in semianalytical models of galaxy formation \citep{DiazGimenez&Mamon10,DiazGimenez+20} and found by \cite{Mamon08} in a compact group previously discovered \citep{Mamon89} inside the Virgo cluster using accurate redshift-independent distances.\footnote{Admittedly, regular groups extracted with the \cite{Yang+07} group finder are also not immune to chance alignments, although probably less so than CG{}s given the latter's permissive accordant velocity criterion.} 
An analysis of five different SAMs indicated that CGs, although selected to be isolated in redshift space, are very rarely isolated in 3D \citep{Taverna+22}. Combining this with their association with the cores of parent groups indicates that many CGs are indeed chance alignments of galaxies within larger groups, usually sharing the BGG.
The \CG{}s that are split between several parent Lim groups are likely to be the most affected by such chance alignments, given their much higher velocity dispersions.

In a forthcoming study, we will analyze the links between compact group properties and their galaxy population. We will also compare the properties of galaxies in compact groups with those within the cores of regular groups.

\begin{acknowledgements}
We thank the anonymous referee for constructive comments after a very
thorough reading of the manuscript.
MT warmly thanks the Institut d’Astrophysique de Paris for its
hospitality.
\end{acknowledgements}

\bibliographystyle{aa}
\bibliography{Paper_AA}

\onecolumn
\appendix
\section{Conversion of virial masses and of virial radii}
\label{sec:conversion}
This appendix provides a simpler, yet more complete formalism than that of appendix~A of \cite{Trevisan+17} to convert virial masses and virial radii, from a prior to final overdensity.

Given that the critical density of the Universe at epoch $z$ is
\begin{equation}
    \rho_{\rm c}(z) =
    \frac{\overline\rho_\mathrm{U}(z)}{\Omega_\mathrm{m}(z)}
    = \frac{3 H^2}{8\pi\,G} = \frac{3 H_0^2}{8\pi\,G}\,E^2(z) \ ,
    \label{rhocrit}
\end{equation}
where 
$\overline\rho_\mathrm{U}(z)$ is the mean density of the Universe at epoch $z$, while
$E^2(z) \equiv [H(z)/H_0]^2 = \Omega_\mathrm{m,0}(1+z)^3 + 1-\Omega_{\mathrm{m},0}$ for a flat Universe,
the mass within the sphere (of radius $r_\Delta$) whose mean density is $\Delta$ times the critical (not mean) density of the Universe is
\begin{equation}
    {\cal M}_\Delta \equiv \mathcal{M}(r_\Delta) = \frac{4\pi}{3}\,\Delta\, r_\Delta^3 \,\rho_c(z) = \left(\frac{\Delta}{2}\right)\,\frac{H_0^2}{G} \,E^2(z)\,r_\Delta^3\ .
    \label{MDelta}
\end{equation}
Comparing now  the mass of the same system within overdensities $\Delta$ and $\Delta'$, Eq.~(\ref{MDelta}) trivially yields
\begin{equation}
    {{\cal M}_{\Delta'} \over {\cal M}_\Delta} = {\Delta'\over \Delta}\,
    \left({r_{\Delta'}\over r_\Delta}\right)^3 \ .
    \label{Mratio1}
\end{equation}
Writing the mass profile as $\mathcal{M}(r) = {\cal M}_\Delta\,\widehat M(r/r_\Delta,r_\Delta/r_{\rm s})$, where $r_s$ is a universal attribute (e.g., the scale radius) of the density profile, while $\widehat M(1,c)=1\ \forall c$,
one trivially obtains (independently of Eq.~[\ref{Mratio1}])
\begin{equation}
      {{\cal M}_{\Delta'} \over {\cal M}_\Delta} = 
      \widehat M \left({r_{\Delta'}\over r_\Delta},{r_\Delta\over r_{\rm s}}\right) \ ,
      \label{Mratio2}
\end{equation}
where $r_\Delta/r_\mathrm{s} \equiv c_\Delta$ is the concentration for the prior overdensity.
Then, eliminating ${{\cal M}_{\Delta'} / {\cal M}_\Delta}$ from Eqs.~(\ref{Mratio1}) and (\ref{Mratio2}) yields
\begin{equation}
    {\Delta'\over \Delta}\,
    \left({r_{\Delta'}\over r_\Delta}\right)^3 = 
    \widehat M \left({r_{\Delta'}\over r_\Delta},{r_\Delta\over r_{\rm s}}\right) \ .
\label{eqtosolve}
\end{equation}
The final radius, $r_{\Delta'}$ is then obtained by numerically solving
Eq.~(\ref{eqtosolve}) for $r_{\Delta'}/r_\Delta$.
Then, the final mass, $\mathcal{M}_{\Delta'}$ is deduced from
 either Eq.~(\ref{Mratio1}) or Eq.~(\ref{Mratio2}).\footnote{Eq.~(\ref{Mratio1}) is simpler than Eq.~(\ref{Mratio2}).
}

The ratios $r_{\Delta'}/r_\Delta$ and $\mathcal{M}_{\Delta'}/\mathcal{M}_\Delta$, calculated with Eqs.~(\ref{eqtosolve}) and (\ref{Mratio1}),
are well approximated by second-order polynomial fits:
\begin{subequations}
\begin{flalign}
    \frac{r_{\Delta'}}{r_{\Delta}} &= \mathrm{dex}\sum_{i=0}^2 a_i \left[\log_{10}\left(\frac{r_\Delta}{r_\mathrm{s}}\right)\right]^i
    \label{coeffs_r} \ ,
    \\
        \frac{\mathcal{M}_{\Delta'}}{\mathcal{M}_{\Delta}} &= \mathrm{dex}\sum_{i=0}^2 b_i \left[\log_{10}\left(\frac{r_{\Delta}}{r_\mathrm{s}}\right)\right]^i \ ,
        \label{coeffs_m}
\end{flalign}
\label{coeffs}
\end{subequations}
\!\!where $r_\mathrm{s}$ is the scale radius of the halo (hence, $r_\Delta/r_\mathrm{s}$ is the concentration of the halo in the prior system), with rms errors  less than 0.0004 dex and 0.0015 dex for radius and mass ratios, respectively.
\begin{table}
\caption{Coefficients of fits to ratios of $z$=0 halo radii and masses.}
    \centering
 \begin{tabular}{llrrrcrrr}
\hline
\hline
 \multicolumn{1}{c}{$\Delta$} & \multicolumn{1}{c}{$\Delta'$} &  \multicolumn{3}{c}{$r_{\Delta'}/r_{\Delta}$}
 & &
 \multicolumn{3}{c}{$\mathcal{M}_{\Delta'}/\mathcal{M}_{\Delta}$}
 \\
 \cline{3-5}
 \cline{7-9}
& &  \multicolumn{1}{c}{$a_0$} & \multicolumn{1}{c}{$a_1$} & \multicolumn{1}{c}{$a_2$} & & 
 \multicolumn{1}{c}{$b_0$} & \multicolumn{1}{c}{$b_1$} & \multicolumn{1}{c}{$b_2$} \\
 \multicolumn{1}{c}{(1)} & \multicolumn{1}{c}{(2)} &\multicolumn{1}{c}{(3)} & \multicolumn{1}{c}{(4)}
 & \multicolumn{1}{c}{(5)} & & \multicolumn{1}{c}{(6)} & \multicolumn{1}{c}{(7)} & \multicolumn{1}{c}{(8)} \\
 \hline
  180m &   200m &     --0.026 &      0.011 &     --0.003 &&     --0.032 &      0.032 &     --0.010 \\
  180m &  vir &     --0.156 &      0.069 &     --0.022 &&     --0.203 &      0.207 &     --0.067 \\
  180m &  200c &     --0.345 &      0.166 &     --0.054 &&     --0.477 &      0.497 &     --0.163 \\
  200m &  vir &     --0.129 &      0.056 &     --0.018 &&     --0.166 &      0.168 &     --0.054 \\
  200m &  200c &     --0.314 &      0.149 &     --0.049 &&     --0.430 &      0.446 &     --0.146 \\
 vir &  200c &     --0.172 &      0.077 &     --0.025 &&     --0.225 &      0.230 &     --0.074 \\
\hline
\end{tabular}
\tablefoot{The ratios of radii and masses are obtained with Eqs.~(\ref{coeffs_r}) and (\ref{coeffs_m}), respectively. The first two columns are
respectively the prior and final overdensities, with suffixes ``m" and ``c" for relative to mean and critical densities of Universe, respectively; ``vir" refers to the \citealt{Bryan&Norman98} virial radius. In terms of the critical density of the Universe, at $z=0$ one has $\Delta_\mathrm{c,180m}= 180\,\Omega_m=55.4$; $\Delta_\mathrm{c,200m}= 200\,\Omega_m=61.6$, while $\Delta_\mathrm{c,vir}=102.2$. Columns
3 to 5 and 6 to 8 are the second-order polynomial coefficients for the ratios of final to prior halo radius and mass, respectively in terms of concentration, following Eqs.~(\ref{coeffs}).}
\label{tab:polyfits}
\end{table}
For future users, we provide the coefficients of the polynomials in Table~\ref{tab:polyfits} for some popular pairs of overdensities, for
the NFW model \citep*{Navarro+96}, for which \citep{Cole&Lacey96}
\begin{equation}
\widehat M(x,c) = \frac{\ln(c x+1)-c x/(c x+1)}{\ln(c+1) - c/(c+1)} \ .
\end{equation}

There are two complications to using the simple Eq.~(\ref{eqtosolve}) for converting virial radii and masses:
1) The group masses are defined using group overdensities relative to the mean density $\rho_\mathrm{U}(z)$ of the Universe, while it is more common to use overdensities relative to the critical density $\rho_\mathrm{c}(z)$ of the Universe. 2)  In this context, Eq.~(\ref{eqtosolve}) is insufficient, because the concentration $c=r_\Delta/r_\mathrm{s}$ in the second term of its right-hand side, defined using the overdensity relative to $\overline\rho_\mathrm{U}$, is usually not as well known as is the concentration defined using the overdensity relative to $\rho_\mathrm{c}$.
Eq.~(\ref{eqtosolve}) can be generalized by expressing the concentration $c_\Delta$ in the original overdensity in terms of the known expression for the concentration $c_\Delta'$ in the new overdensity:
\begin{equation}
c_\Delta = \frac{r_\Delta}{r_\mathrm{s}} = \frac{r_\Delta}{r_\Delta'}\,c_\Delta'\left(\mathcal{M}_\Delta\,\frac{\mathcal{M}_\Delta'}{\mathcal{M}_\Delta}\right) 
= \frac{r_\Delta}{r_\Delta'}\,c_\Delta'\left[\mathcal{M}_\Delta\,\frac{\Delta'}{\Delta}\,\left(\frac{r_\Delta'}{r_\Delta}\right)^3\right]
\ ,
\label{cofMDelta}
\end{equation}
where the second equality is obtained using Eq.~(\ref{Mratio1}).
Inserting Eq.~(\ref{cofMDelta}) into Eq.~(\ref{eqtosolve}) yields
\begin{equation}
       {\Delta'\over \Delta}\,
    \left({r_{\Delta'}\over r_\Delta}\right)^3 = 
    \widehat M \left[{r_{\Delta'}\over r_\Delta},
    \frac{r_\Delta}{r_\Delta'}\,c_\Delta'\left(\mathcal{M}_\Delta\,\frac{\Delta'}{\Delta}\,\left(\frac{r_\Delta'}{r_\Delta}\right)^3\right)\right] \ ,
    \label{eqtosolvegen}
\end{equation} 
which can be numerically solved for $r_\Delta'/r_\Delta$, knowing the ratio of overdensities, $\Delta'/\Delta$, and the mass for the initial overdensity, $\mathcal{M}_\Delta$. The mass ratio is then deduced from 
Eq.~(\ref{Mratio1}) and can be checked using Eq.~(\ref{Mratio2}).

We now discuss how  we converted the masses $\mathcal{M}_\mathrm{180m}$ to $\mathcal{M}_\mathrm{200c}$ in Sect.~\ref{sec:CGBGG_in_PC}.
Analogous to Eq.~(\ref{MDelta}), the mass within the radius of mean density equal to $\Delta_{\rm m}$ times the mean density of the Universe is
\begin{equation}
    \mathcal{M}_{\Delta,\mathrm{m}} \equiv {\cal M}(r_{\rm \Delta_{\rm m}}) =\frac{4\pi}{3}\,\Delta_\mathrm{m}\,\Omega_\mathrm{m}(z)\,\rho_\mathrm{c}(z)\,r_{\Delta,\mathrm{m}}^3  = \frac{\Delta_\mathrm{m}}{2}\,\frac{\Omega_\mathrm{m,0}\,H_0^2}{G}\,(1+z)^3\,r_{\Delta,\mathrm{m}}^3  \ .
    \label{MDeltam}
\end{equation}
Then, by comparing the second equalities of Eqs.~(\ref{MDelta}) and (\ref{MDeltam}), the ratio of overdensity relative to $\rho_\mathrm{c}(z)$ to overdensity relative to $\overline \rho_\mathrm{U}(z)$ is 
\begin{equation}
    \frac{\Delta}{\Delta_\mathrm{m}} = \frac{\Omega_{\mathrm{m},0}\,(1+z)^3} {E^2(z)}\ . 
    \label{overdenscritovermean}
\end{equation}
We note that combining Eqs.~(\ref{Mratio1}) and (\ref{overdenscritovermean}),
leads to a mass ratio:
\begin{equation}
    \frac{\mathcal{M}_{\Delta'}}{\mathcal{M}_{\Delta,\mathrm{m}}} = \frac{\Delta'}{\Omega_{\mathrm{m},0}\,\Delta_\mathrm{m}}
    \,\frac{E^2(z)}{(1+z)^3}\,
    \left(
    \frac{r_{\Delta'}}{r_{\Delta,\mathrm{m}}} \right)^3\ .
    \label{MDelta2overMDeltam}
\end{equation}
Inserting Eq.~(\ref{overdenscritovermean}) into Eq.~(\ref{eqtosolvegen}), yields
\begin{equation}
     \frac{\Delta'}{\Omega_{\mathrm{m},0}\,\Delta_\mathrm{m}}
    \,\frac{E^2(z)}{(1+z)^3}\,
    \left(
    \frac{r_{\Delta'}}{r_{\Delta,\mathrm{m}}} \right)^3
    =\widehat M \left\{{r_{\Delta'}\over r_\Delta},
    {r_\Delta\over r_\Delta'}\,
    c\left[\frac{\Delta'/\Delta_\mathrm{m}}{\Omega_{\mathrm{m},0}}
    \,\frac{E^2(z)}{(1+z)^3}\,
    \left(
    \frac{r_{\Delta'}}{r_{\Delta,\mathrm{m}}} \right)^3 \,\mathcal{M}_{\Delta,\mathrm{m}}\right]\right\}\ ,
    \label{Ratio2}
\end{equation}
which can be numerically solved for $r_{\Delta'}/r_{\Delta,\mathrm{m}}$.
The mass ratio is then deduced using Eqs.~(\ref{Mratio1}) and (\ref{overdenscritovermean}) and can be verified using Eq.~(\ref{Mratio2}).

Hence, $\Delta = 180\,\Omega_\mathrm{m} \simeq 55.44$ and $\Delta' = 200$. We adopted the $z$=0 concentration-mass relation of \cite{Dutton&Maccio14} for $\Delta'=200\mathrm{c}$, i.e. 
\begin{equation}
    c_{200\mathrm{c}} = 8.0\,\mathcal{M}_{200\mathrm{c}}^{-0.101} \ .
    \label{cDM14}
\end{equation} 
Inserting $\Delta = 55.44$ and $\Delta'=200$ in Eq.~(\ref{Ratio2}), we ended up solving
\begin{equation}
3.61\left(\frac{r_\mathrm{200c}}{r_\mathrm{180m}}\right)^3 
= \widehat M\left\{
\frac{r_\mathrm{200c}}{r_\mathrm{180m}},
\frac{1}{3.61}\,c_\mathrm{200c}\left[3.61\left(\frac{r_\mathrm{200c}}{r_\mathrm{180m}}\right)^3
\mathcal{M}_\mathrm{180m}\right]
\right\} 
\label{eqtosolvenum}
\end{equation}
for $r_\mathrm{200c}/r_\mathrm{180m}$ (where 3.61=200/55.44).
Numerically solving Eq.~(\ref{eqtosolvenum}), we found the following power-law approximations:
\begin{subequations}
    \begin{flalign}
         \frac{r_\mathrm{200c}}{r_\mathrm{180m}} & \simeq 0.603\,\left(\frac{h\,r_\mathrm{180m}}{100\,\mathrm{kpc}}\right)^{-0.0156}  \ ,
         \label{r_ratio_fit} \\
        \frac{\mathcal{M}_\mathrm{200c}}{\mathcal{M}_\mathrm{180m}} & \simeq 0.756\,\left(\frac{h\,\mathcal{M}_\mathrm{180m}}{10^{12}\mathrm{M_\odot}}\right)^{-0.0156} \ ,
        \label{M_ratio_fit} 
    \end{flalign}
\end{subequations}
for our choice of the $z$=0 \cite{Dutton&Maccio14} concentration-mass relation (Eq.~\ref{cDM14}), as well as $\Omega_\mathrm{m}=0.308$. 
The very shallow power-laws are a consequence of the shallow concentration-mass relation (Eq.~[\ref{cDM14}]) and the logarithmic increase of NFW mass with radius in the envelope ($r_{180\mathrm{m}} > r_{200\mathrm{c}} >  3\,r_\mathrm{s}$).
Eqs.~(\ref{r_ratio_fit}) and (\ref{M_ratio_fit}) are both precise to 0.2\% rms for $12 < \log (h\,\mathcal{M}_\mathrm{180m}/\mathrm{M}_\odot) < 15$.
Slight changes in the cosmological parameters should only have a very small effect.

\label{lastpage}

\end{document}